\documentclass[11pt]{report}%
\usepackage{setspace}
\usepackage{graphicx}
\usepackage{epstopdf}
\usepackage{amsmath}%
\usepackage{amsfonts}%
\usepackage{amssymb}%
\usepackage{epsf}%
\usepackage{hyperref}%
\usepackage{eso-pic}
\usepackage{graphicx}
\AddToShipoutPicture*{%
\put(0,0){\includegraphics[width=\paperwidth]{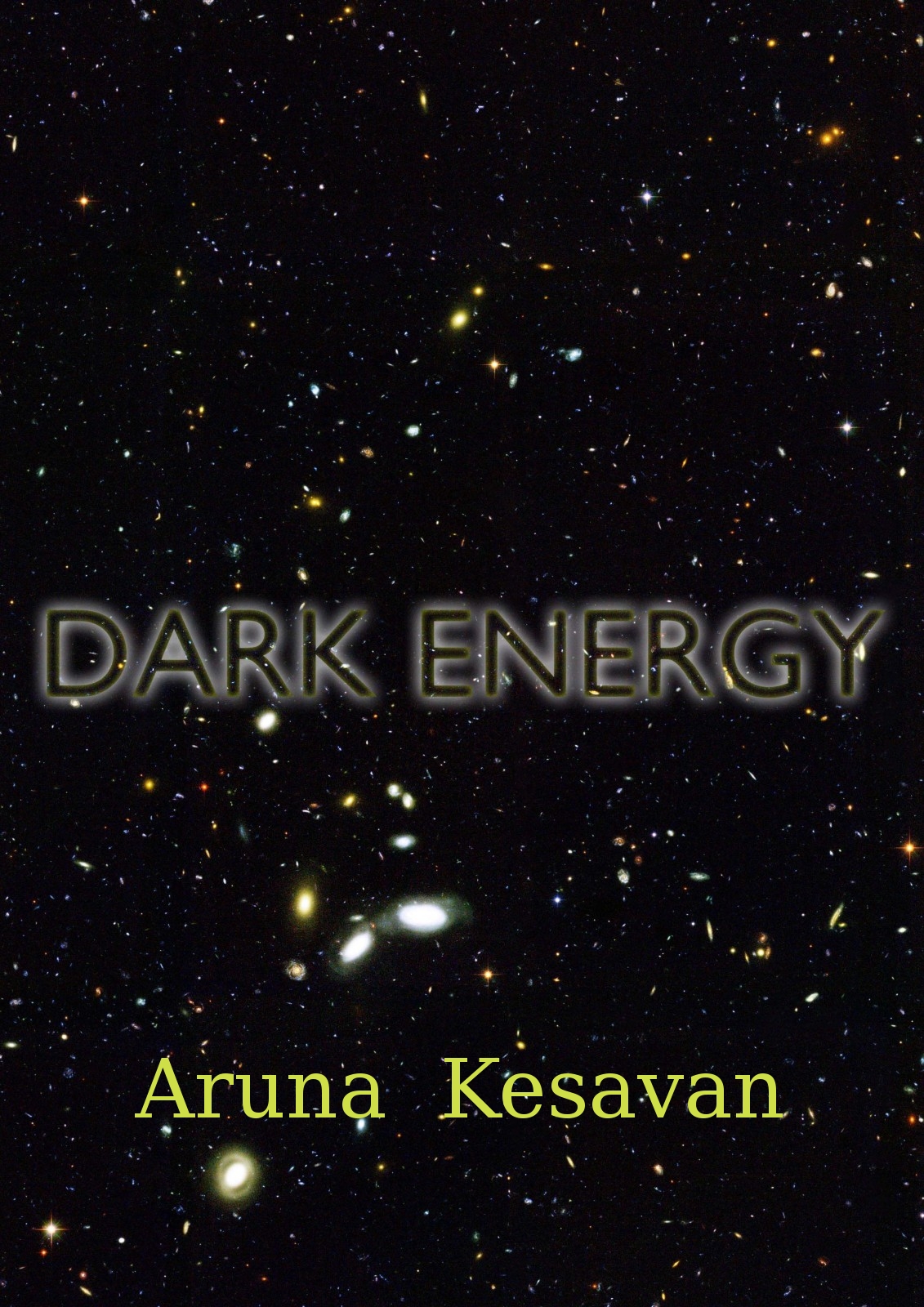}}%
}

\def\singlespace {\smallskipamount=3.75pt plus1pt minus1pt
                  \medskipamount=7.5pt plus2pt minus2pt
                  \bigskipamount=15pt plus4pt minus4pt
                  \normalbaselineskip=15pt plus0pt minus0pt
                  \normallineskip=1pt
                  \normallineskiplimit=0pt
                  \jot=3.75pt
                  {\def\smallskip {\vskip\smallskipamount}}
                  {\def\medskip   {\vskip\medskipamount}}
                  {\def\bigskip   {\vskip\bigskipamount}}
                  {\setbox\strutbox=\hbox{\vrule
                    height10.5pt depth4.5pt width 0pt}}
                  \parskip 7.5pt
                  \normalbaselines}
\def\middlespace {\smallskipamount=5.625pt plus1.5pt minus1.5pt
                  \medskipamount=11.25pt plus3pt minus3pt
                  \bigskipamount=22.5pt plus6pt minus6pt
                  \normalbaselineskip=22.5pt plus0pt minus0pt
                  \normallineskip=1pt
                  \normallineskiplimit=0pt
                  \jot=5.625pt
                  {\def\smallskip {\vskip\smallskipamount}}
                  {\def\medskip   {\vskip\medskipamount}}
                  {\def\bigskip   {\vskip\bigskipamount}}
                  {\setbox\strutbox=\hbox{\vrule
                    height15.75pt depth6.75pt width 0pt}}
                  \parskip 11.25pt
                  \normalbaselines}
\def\doublespace {\smallskipamount=7.5pt plus2pt minus2pt
                  \medskipamount=15pt plus4pt minus4pt
                  \bigskipamount=30pt plus8pt minus8pt
                  \normalbaselineskip=30pt plus0pt minus0pt
                  \normallineskip=2pt
                  \normallineskiplimit=0pt
                  \jot=7.5pt
                  {\def\smallskip {\vskip\smallskipamount}}
                  {\def\medskip   {\vskip\medskipamount}}
                  {\def\bigskip   {\vskip\bigskipamount}}
                  {\setbox\strutbox=\hbox{\vrule
                    height21.0pt depth9.0pt width 0pt}}
                  \parskip 15.0pt
                  \normalbaselines}
\makeatletter

\newcommand{\Rmnum}[1]{\expandafter\@slowromancap\romannumeral #1@}
\makeatother

\begin{document}

\newpage
\thispagestyle{empty}
\begin{center}
\includegraphics[width=0mm]{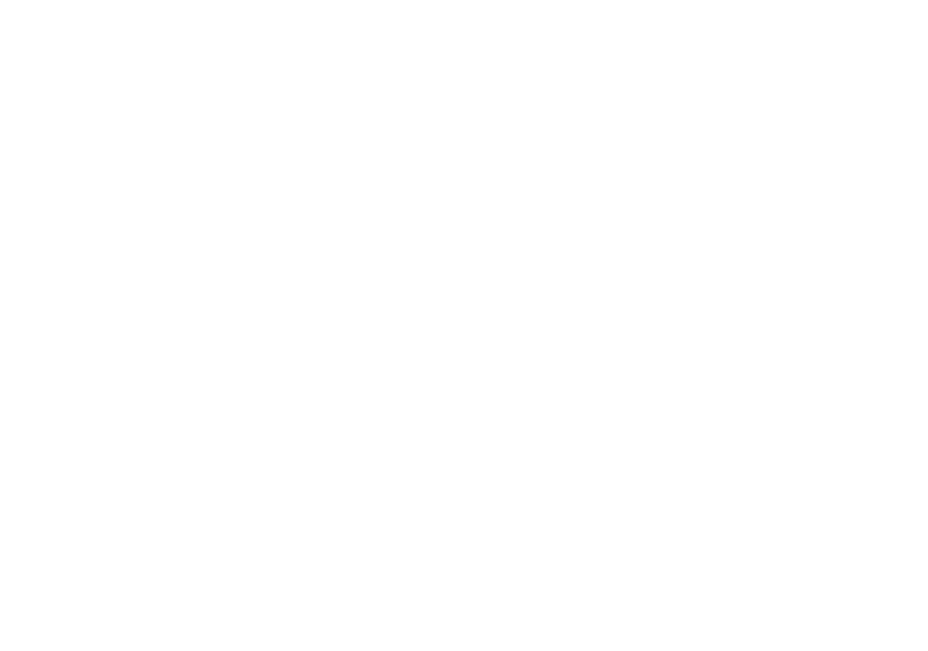}
\end{center}

\newpage
\begin{center}
\thispagestyle{empty}
{\bf {\Huge DARK ENERGY}}\\

\bigskip
\bigskip
\bigskip
\bigskip
\bigskip
\bigskip

Thesis\\
Submitted in partial fulfillment of the requirements of\\
BITS C421T/422T Thesis\\

\bigskip
\bigskip

By\\

\bigskip
\bigskip

{\bf {\Large Aruna Kesavan}}\footnote{aruna.kesavan@gmail.com}\\
ID No. 2004B5B4008\\

\bigskip
\bigskip

Under the supervision of\\

\bigskip
\bigskip
\bigskip

{\bf {\Large Joseph Samuel}}\\
Professor, Theoretical Physics Department\\
Raman Research Institute, Bangalore - 560080\\

\vspace{1 cm}
\includegraphics[width=20mm]{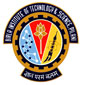}

BIRLA INSTITUTE OF TECHNOLOGY AND SCIENCE, PILANI (RAJASTHAN)\\
\bigskip

May 25, 2009
\end{center}

\newpage
\thispagestyle{empty}
\begin{center}
{\bf {\Large {CERTIFICATE}}}\\
\end{center}
\begin{figure}[here]
\centering
\includegraphics[scale=0.30]{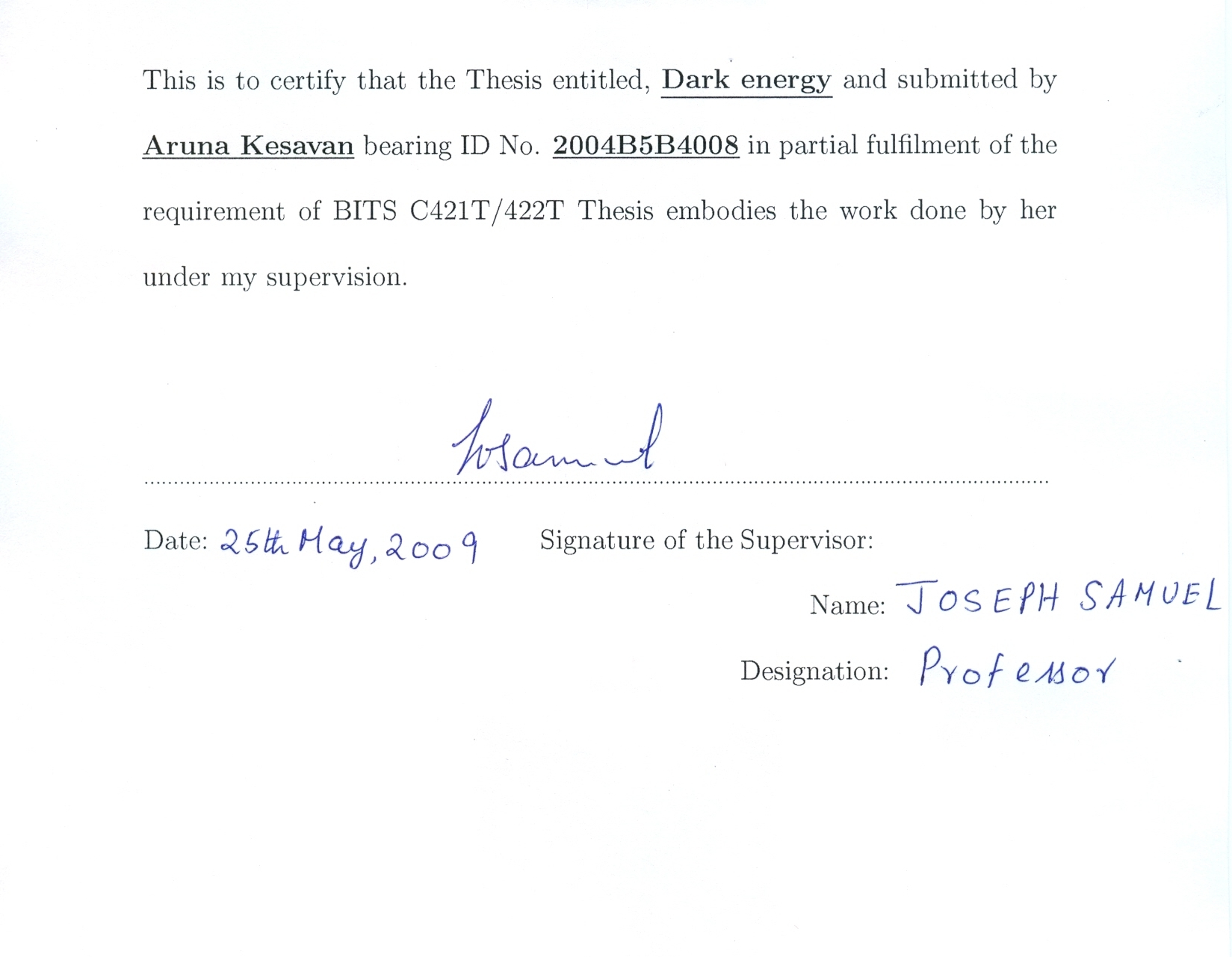}
\end{figure}
\setcounter{page}{0}

\newpage
\thispagestyle{empty}
\setcounter{page}{0}
\begin{center}
\section*{\bf {\Large {ACKNOWLEDGMENTS}}}
\end{center}
\vspace{0.5 cm}
\doublespacing
{\Large
This work owes a great deal to a great many people.\\

I am grateful to my mentor, Sam, for being a great guide. He has been patient, clever, inspiring and a lot of fun. His lucid explanations and apt analogies have many a time helped me make sense of obscure ideas. Most of all, I want to thank him for helping me have fun while chasing my questions, both within Physics and outside. I also wish to acknowledge the important part he played in my getting this done on time!\\

My sincere thanks are due to everyone who attended the talks I gave on the subject during the course of this work. They include Sam, Supurna, Poonam, Sreedhar, Simi, Rohit, Sneha, Anupam, Giri, Anagha, Pragya and Kalpana. The discussions we had have been immensely useful in furthering my own understanding.\\

Special thanks to Dr. Lakshmi Saripalli for reading the report and making important suggestions.\\

I thank Jeremy Mayes for allowing me to modify and use his cartoon `Isaac Newton \& apple' in the back cover of this report.

\newpage
\thispagestyle{empty}
\setcounter{page}{0}
{\Large
I am obliged to RRI for funding and accommodating me under the Visiting Students Programme.\\

Many thanks to all the staff at RRI for sorting out problems efficiently and helping me use the facilities extended to me, and hence, making my stay here very comfortable.\\}

I wish to thank Preethi, Simi, Sneha, Kalpana, Nabaneeta, Sreekanth, Aasif, Mohesh, Uma, Rohit and the others for all the good times.\\

My stay in Bangalore would have been very dull had it not been for my many relatives and friends. I wish to thank them for all the rejuvenating weekends. My special thanks to Lakshmi and Jan.\\

Finally, I cannot thank Ma, Daddy, Veena and Bhargav enough for their love, support and everything else.
}
\setcounter{page}{0}

\newpage
\thispagestyle{empty}
\setcounter{page}{0}
\begin{center}
\section*{\bf \Large{ABSTRACT}}
\end{center}
\large{
Dark energy is one of the mysteries of modern science. It is unlike any known form of matter or energy and has been detected so far only by its gravitational effect of repulsion. Owing to its effects being discernible only at very very large distance scales, dark energy was only detected at the turn of the last century when technology had advanced enough to observe a greater part of the universe in finer detail. The aim of the report is to gain a better understanding of the mysterious dark energy. To this end, both theoretical methods and observational evidence are studied. Three lines of evidence, namely, the redshift data of type \Rmnum{1}a supernovae, estimates of the age of the universe by various methods, and the anisotropies in the cosmic background radiation, build the case for existence of dark energy. The supernova data indicate that the expansion of the universe is accelerating. The ages of the oldest star clusters in the universe indicate that the universe is older than previously thought to be. The anisotropies in the cosmic microwave background radiation suggest that the universe is globally spatially flat. If one agrees that the dynamics of the geometry of the universe is dictated by its energy-momentum content through Einstein's general theory of relativity, then all these independent observations lead to the amazing conclusion that the amount of energy in the universe that is presently accounted for by matter and radiation is not enough to explain these phenomena. One of the best and simplest explanations for dark energy is the cosmological constant. While it does not answer all questions, it certainly does manage to explain the observations. The following report examines in some detail the dark energy problem and the candidacy of the cosmological constant as the right theory of dark energy.
}

\setcounter{page}{0}
\thispagestyle{empty}
\addtocontents{toc}{\protect\thispagestyle{empty}}
\tableofcontents
\setcounter{page}{0}
\thispagestyle{empty}

\newpage
\pagestyle{plain}
\setcounter{page}{1}

\chapter{INTRODUCTION}\label{chap1}The heavens have always bred wonder and curiosity in human beings. A few million years since the first human being looked up at the sky and wondered about it, the universe continues to challenge our imagination. The universe is the ultimate jigsaw puzzle. And the last decade has brought to light some interesting new pieces of the puzzle. Scientists are busy asking questions, the answers to which will help us determine how to fit these pieces together to form a larger picture. In this report, we ask and seek to answer two simple questions about the new pieces of the puzzle. \\{\it What is dark energy? Why is dark energy believed to exist?} \\The answers we seek are not easy or straightforward. In most cases the attempt at an answer only leads to new questions. To make a useful study, we concentrate our efforts in this report on the simplest candidate for dark energy, the cosmological constant. To get a feel for the problem of dark energy, we consider the following simple scenario. We try to understand in terms of known Physics what happens to a stone of unit mass that is thrown vertically upwards from the surface of the earth. The total energy of the stone is the sum of its gravitational and potential energies, given by Newton's laws as $E = \frac{1}{2}\dot{r}^2 - \frac{G\,M}{r}$, where $r$ is the radial distance of the stone from the centre of mass of the earth and $M$ is the total mass of the earth. The negative sign of the potential energy comes from the attractive nature of the force of gravitation between the earth and the stone. From Newton's law we obtain the following expression for the acceleration of the stone as a function of $r$.\begin{equation}\ddot{r} = -\frac{G\,M}{r^2}\label{stoneacce}\end{equation}\begin{figure}\centering\includegraphics[width=150mm]{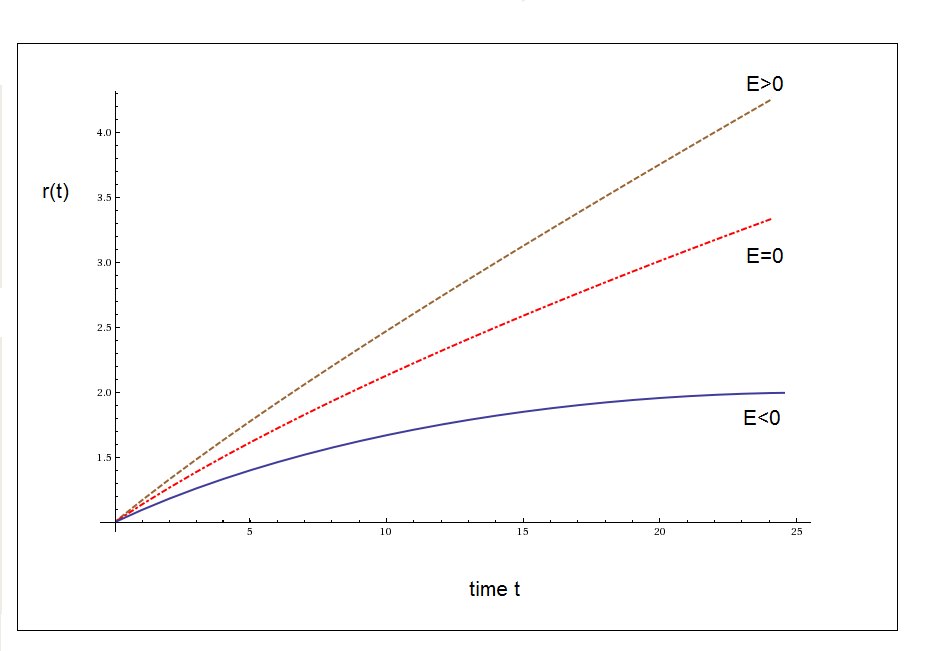}\caption{\label{stone} Evolution of distance of the stone from the earth for different values of its total energy.}\end{figure}As expected this implies that the total energy of the particle is a constant of its motion. The attractive force of gravity tends to slow the stone down. If the total energy is negative, it brings the stone to a momentary stop and reverse its direction of motion at a finite radial distance. For zero and positive total energies, although the stone is slowed down, it is never brought to a halt at a finite distance from the earth. In the two cases, the stone has just and more than enough kinetic energies respectively to overcome gravity and escape from the earth, never to return. In particular, for zero total energy, $r \propto t^{2/3}$. If the $r$ coordinate of the stone is plotted against time, zero acceleration will correspond to a straight line of constant slope that is the speed of the stone. All negatively accelerated curves fall below the slope at each point on the curve, while those with positive acceleration are above the slope at each point on the curve. Newtonian gravity predicts the existence of trajectories of only zero or negative acceleration. Zero acceleration is achieved when $M=0$ that is when there is no earth to exert a force on the stone and the stone coasts along with $r \propto t$. Here $r$ is the distance from some origin of the coordinate system. Generalising our example from two particles (the earth and stone) to a cloud of particles that are in free fall, the result is that the second derivative of the volume of the cloud must be negative. This result is also true in the general theory of relativity in the form that Einstein proposed it in 1915. But observations in the last decade have challenged ideas that we have held true for long and demand explanation. The universe, against all expectation, is found to be expanding with a positive acceleration! This observation is like watching our stone go faster as it goes farther away from the earth, though, at a much larger scale. Attractive gravity cannot explain this phenomenon. Whatever it is that causes the acceleration is termed dark energy. We give such a broad definition of dark energy here, and talk of candidates for dark energy, because there is, as yet, no consensus among scientists about how to explain the observed acceleration. Most of present day cosmology is based on what is known as the cosmological principle, the notion that ours is no special position in the universe, and that the universe appears homogeneous and isotropic about every point. Some believe that the observed acceleration is the effect of inhomogeneities in the universe and that there is no need to invoke the idea of dark energy at all. One just needs to reconsider the cosmological principle. Some others believe that the accelerated expansion is owing to a form of energy with some equation of state, which is again, not universally agreed upon. There is also varying opinion about whether or not the equation of state of dark energy is time-dependent. In short, the questions thrown up by the astounding discoveries of the recent past are far from having been satisfactorily answered now. Apart from accelerated expansion, there are other lines of evidence that point towards the existence of dark energy. In this report, we attempt to understand some of these, while studying the simplest candidate for dark energy, the cosmological constant. The cosmological constant is a term added to Einstein's equations in his theory of gravity, the general theory of relativity, that can introduce the repulsive effects in gravity that can explain the accelerated expansion. The cosmological constant is a strong contender for the right theory of dark energy as not only does it account for the observations, it can also be given a physical interpretation as the Lorentz-invariant energy of vacuum. There is, however, a problem, in that the size of the cosmological constant differs widely from the value of vacuum energy predicted by physical theories.This report is divided into eight chapters and has three appendices attached. The first chapter after the introduction here tracks the story of the cosmological constant from its birth as an attempt to salvage a static universe to its going out of fashion when the expansion of the universe was detected. In the chapter after that, we get a glimpse of the great potential of the general theory of relativity as a dynamic theory of gravity. We describe an expanding universe and study how various forms of energy affect the evolution of the universe. We will familiarise ourselves with theoretical tools that will later help us make sense of the recent observations. In the next part of the thesis we shift our focus on the startling observations of the past decade. We start off with evidence for the accelerated expansion of the universe that comes from studying the luminosity distance-redshift relation for distant type Ia supernovae. We then move on to discuss limits on the global curvature determined from observations of anisotropies in the cosmic background radiation and lower limit on the age of the universe from the ages of the oldest star clusters. All of these strongly support the existence of dark energy and seem to concur on the amount of dark energy present in the universe today. The astonishing conclusion from these and other observations seems to be that almost $96 \%$ of the energy content of the universe is in the form of dark content, that is, either dark energy (about $72 \%$) or dark matter (about $24 \%$). Not much mention is made of dark matter in this report. Wherever densities of matter are quoted, it includes contributions from both ordinary and dark matter. Although the focus of this report is not on dark matter, it is good to mention a few things about dark matter so as to avoid any confusion between the two dark components of the universe. The main evidence for the existence of dark matter comes from analysing the orbital velocities of stars in galaxies. As the velocities of stars at a certain distance $r$ from the centre of the galaxy is determined by the amount of matter contained within a sphere of radius $r$ centred at the galaxy centre, it is possible to estimate the amount of matter in galaxies. Another independent estimate of the mass contained in galaxies is the total luminosity of the galaxy. Data collected from many galaxies suggest that these two estimates to do match up. There seems to be much more matter in the galaxies than is luminous. This unseen matter has been termed dark matter. Additionally, this dark matter is assumed to be cold dark matter as it is not observed to radiate, and its pressure is taken to be negligible. What dark matter does share in common with dark energy is that both of them have been detected only through their gravitational effects. As we shall see, dark energy has properties very different from any known form of energy, including dark matter, and these properties are essential in order to explain such observations as the accelerated expansion of the universe.In the penultimate chapter of this report we examine the interpretation of the cosmological constant as vacuum energy. In the concluding chapter we make remarks about the many questions that have grown out of our original two questions. We also broaden our view a little to mention what other ideas scientists have about dark energy. The appendices include some topics of interest to the author that are not directly related to the problems posed by dark energy, but which nevertheless turned up during the study of the problems. These include an analysis of Olbers' paradox, a study of symmetric spaces and Killing vectors and a brief account of various distance estimation methods used in astronomy.Well, let us begin! 

\chapter{THE COSMOLOGICAL CONSTANT}
\label{chap2}

The history of science lacks not in drama, for after all, science has always been about beliefs and their refutation or vindication. The story below traces a small part of the evolution of our beliefs about the nature of the universe we live in. Our story begins with Albert Einstein and his belief. Like many great thinkers before his time he considered the universe to be static, eternally unchanging. Upon writing down the equations of the general theory of relativity, he realised that these were not consistent with the premise of a static universe. He chose to alter his equations in such a way that they allowed for the static character of the universe instead of acceding that the universe evolved with time. This led him to introduce a term, later called the cosmological constant, into his equations. However, soon Hubble's observations of galaxy redshifts forced the abandonment of the idea of a static universe. And for many decades after that it was believed that the cosmological constant had to be set to zero. That was until the startling observations of the late 1990's. These will, however, be dealt with in another chapter.
In the current chapter, we discuss the details of Einstein's equations applied to a static homogeneous, isotropic space-time metric of the universe.

We start by elaborating on the assumptions cosmologists make about the universe.
In general, studying simple systems at first is both useful and easier than tackling complex systems, and symmetries simplify systems. The particular symmetries of homogeneity and isotropy of the universe, that cosmologists these days believe in, reflect the rejection of a special spacetime position for the earth and the observers on it, a trend that began in modern science with Copernicus stating and proving that the earth orbited the sun and hence, was not the centre of the universe. Over the years, observations probing deeper into space have only justified this point of view as the universe seems uniform in large scales of hundreds of millions of light years. Our sun is just another typical star in just another typical galaxy in just another typical galaxy cluster out of many. Furthermore, the highly isotropic background radiation \footnote{To be discussed in more detail in Chapter \ref{chap5}.} that has been observed to permeate space around us gives us reason to believe our assumptions are correct.
By a static universe one means that it has not been expanding or contracting and by eternal one means that the universe has no beginning or end. That the universe was static and eternal was easy enough to believe in Einstein's time because centuries of astronomical observations seemed to indicate that the universe on a large scale had not changed in all that time, and hence, there was no reason to believe it might have before human beings came to record its history. The velocities of stars observed were also too small and random to believe that the universe was not static on a large scale. \footnote{Justified as these assumptions were at the time they were believed, they posed a serious problem in understanding why the night sky was dark if the universe was filled with stars. This is called Olbers' paradox and is discussed in Appendix \ref{appa}.}

Translating these assumptions to mathematics, \footnote{A study of symmetric spaces is made in Appendix \ref{appb}.} one finds that the line element of the most general metric space \footnote{The line element is chosen such that the metric has Lorentz signature (+,-,-,-).} with the mentioned properties is given in spherical polar coordinates by
\begin{equation}
d\tau^2 = c^2\,dt^2 - a^2[\frac{dr^2}{1 - K \, r^2} + r^2(d\theta^2 + sin^2 \theta d\phi^2)].
\end{equation}
where the scale factor $a$ and the the curvature of space $K$ are both constants independent of the coordinates $(t, r, \theta, \phi)$.

Einstein's original equations for the gravitational field came from requiring that equations of motion were generally covariant under coordinate transformations and reduced to the Newtonian form in very weak gravitational fields. These related the Ricci tensor, that was made up of second derivatives of the metric tensor, the curvature scalar formed by contracting the Ricci tensor and the energy-momentum content of the universe in the following manner.
\begin{equation}
R_{\mu\nu} - \frac{1}{2} R g_{\mu\nu} = -8 \pi\, G \,T_{\mu\nu}
\label{eineqn}
\end{equation} 

The non-zero components of the Ricci tensor and the curvature scalar are $R_{11}=-2K/(1-Kr^2)$, $R_{22}=-2K r^2$, $R_{33}=-2K r^2 sin^2\theta$ and $R=6K/a^2$ respectively.
Einstein's equations, when applied to the most general static metric yield the following equations if the energy-momentum tensor is taken to be of the form $T^{\mu}_{\nu}= Diagonal(\rho, -p, -p, -p)$ as required in a homogeneous and isotropic universe.

\begin{equation}
K = \frac{8 \pi G \, \rho \, a^2}{3}
\label{stat1}
\end{equation}

\begin{equation}
-K = 8 \pi G \, p \, a^2
\label{stat2}
\end{equation}

If energy density is positive, according to  (\ref{stat1}), the universe must be positively curved to allow for positive $a^2$. Then (\ref{stat2}) implies that pressure must be negative. But for all known forms of energy, pressure is non-negative.  Thus, the above equations can yield no consistent solution for the scale factor, $a$, of the universe. Einstein discovered that he could modify his equations if, in addition to the conditions of general covariance and reduction to the Newtonian limit, he allowed for the inclusion of derivatives of the metric tensor of lower than second order. Since, in a locally inertial frame of reference, the first derivative of the metric tensor vanishes, there are no tensors formed from the first derivatives of the metric tensor. This leaves only the metric tensor itself to be included. The modified equations are given below.

\begin{equation}
R_{\mu\nu} - \frac{1}{2}\, R\, g_{\mu\nu} + \lambda\, g_{\mu\nu} = - 8 \pi G\, T_{\mu\nu}
\label{leineqn}
\end{equation} 

$\lambda$ is called the cosmological constant. It should be noted that the effect of including $\lambda$ in the equations can be observed more prominently in large distance scales where the contributions from higher order derivatives of the metric tensor tend to fall. The modified forms of (\ref{stat1}) and (\ref{stat2}) are:

\begin{equation}
K = \frac{8\, \pi\, G\, \rho a^2}{3} + \frac{\lambda a^2}{3}
\label{lstat1}
\end{equation}

\begin{equation}
-K = 8 \,\pi\, G \, p\, a^2 - \lambda a^2
\label{lstat2}
\end{equation}

For a spherical universe with $K=1$ that is filled with pressureless matter (also called dust) one is able to find a solution for $a$. From (\ref{lstat2}) $\frac{1}{a^2}=\lambda$. Substituting this in (\ref{lstat1}), one obtains that $\lambda=4 \pi G \rho$ .

\section{Hubble's observations and its consequences}
In 1929, Hubble \cite{hub} claimed that the velocities of recession of luminous bodies he had observed were proportional to their distances from the earth. This claim came as a blow to believers in a static universe, for if observers everywhere in the universe noted a linear increase in recessional velocity with distance, it meant that the universe was expanding.
With the establishment of cosmic expansion, the original motivation for the introduction of the cosmological constant was lost. The scale factor of the universe was allowed to evolve with time and $\lambda$ was set to zero.

The metric thus became $d\tau^2 = c^2 \, dt^2 - a^2(t)[dr^2/(1-K\,r^2) + r^2\,(d\theta^2 + sin^2 \theta d\phi^2)]$. This is called the Robertson-Walker metric. An important and useful property of the coordinate system used to write out the line element of the metric space is that it is comoving. To understand what a comoving coordinate system is, let us consider a dense cloud of particles that are in free fall. Let us imagine that each particle in this cloud is associated with a clock that measures time and the time measured by its clock is the particle's time coordinate. Each particle is also associated with some unique spatial coordinate. A comoving coordinate system in this cloud is one in which the spatial coordinates of the particles do not change. From the Robertson-Walker metric we can deduce that a particle at rest in these coordinates will remain at rest as $\Gamma^i_{00}=0$ for $i=1,2,3$.
Thus in a comoving coordinate system, the clocks attached to the particles, in fact, show proper time as the particles follow their respective geodesics. It is useful to synchronise the clocks of all the various particles in such a way that they begin simultaneously at some time when all the particles were at rest relative to each other. Then, if the gravitational field over the cloud were uniform, all the clocks would show the same time. In the description of the whole universe, for instance, the Big Bang can be thought of as the origin of time. The Big Bang will be discussed in the following chapter on the dynamics of the universe. For now it suffices to know that there exists a comoving coordinate system in which we can describe our universe. It is, of course, essential that the geodesics of the particles be such that no two of them intersect. This requirement is the formal statement of Weyl's postulate. Together with the Cosmological Principle, it forms the cornerstone of modern day cosmology.

The Robertson-Walker metric can be rewritten as 
\begin{equation}
d\tau^2 = c^2 \, dt^2 - a^2(t)\,[d\chi^2 + f^2_K(\chi)(d\theta^2 + sin^2 \theta d\phi^2)]
\label{rwmetric}
\end{equation}
where 
\begin{equation*}
f_K(\chi)=
\begin{cases} sin(\chi) & \text{if $K>0$ (spherical space),} \\
	      \chi      & \text{if $K=0$ (flat space),} \\
	sinh(\chi)  & \text{if $K<0$ (hyperbolic space).} \\
\end{cases}
\end{equation*}

Einstein's equations yield the following relations for the evolution of the scale factor and the density of the various energy sources in the universe.
\begin{equation}
\dot{a}^2 + K = \frac{8 \pi G \rho a^2}{3}
\label{dyn1}
\end{equation}
\begin{equation}
\dot{\rho} = - \frac{3 \dot{a}}{a} (\rho + p)
\label{dyn2}
\end{equation}
A detailed discussion of the solutions of these equations is contained in Chapter 3. Presently we focus on understanding Hubble's observations in terms of the Robertson-Walker metric.
The radial velocity, $v$, of a radiating object is estimated by studying its radiation spectrum. A spectrum looks pretty much like a bar code and its characteristic features can be identified even when the wavelengths are redshifted. The positions and relative intensities of absorption and emission lines are studied. These patterns are then compared with spectra of elements and compounds on earth to compute by what amount the wavelengths have been shifted.
If this shift is assumed to be a Doppler shift it is given by $1+z=v/c$. So, Hubble's diagram is actually a graph of the fractional shift in wavelength versus the distance to the radiating object. Hubble calculated the distances to the objects he observed, stars called Cepheid variables, by comparing their absolute and apparent luminosities. Hubble observed that the farther away an object was, the greater was its spectral redshift. In a dynamic universe, this redshift can be seen as being caused by the increase in the scale factor of the universe in the time that a photon takes to travel from the source to the observer. This results in increasing distance between two points and hence, larger wavelengths for photons. Such a redshift is called cosmological redshift, and it is easy to relate the wavelengths at times $t_1$ of emission and $t_0$ of reception of a photon to the ratio of the scale factor of the universe at those times. Without loss of generality, the spatial coordinates of emission in comoving spherical polar coordinates can be taken as ($r_1$,0,0) and those of reception as (0,0,0). Photons follow null geodesics. For a null geodesic, ${c \, dt}^2 = a^2(t)\,d\chi^2$. The gives for the radial coordinate the expression $\int_{t_1}^{t_0} \, c \, dt/a(t) = -\int_{\chi_1}^{0} \, d\chi$. The negative sign is due to decrease in the radial coordinate of photon with increase in time coordinate.
A similar relation can be written for a photon emitted at $t_1 + \delta t_1$ and received at $t_0 + \delta t_0$. Since, in a comoving coordinate system, the coordinates of the emitter and receiver do not change with time, the integrals over time in both equations can be equated to give 
\begin{equation}
\int_{t_1}^{t_1 + \delta t_1} \, {\frac{dt}{a(t)}} = \int_{t_0}^{t_0 + \delta t_0} \, {\frac{dt}{a(t)}}
\end{equation}
Assuming that the scale factor is almost constant in the small intervals of time $\delta t_1$ and $\delta t_0$, the redshift relation is obtained as 
\begin{eqnarray}
1+z &=& \frac{\delta t_0}{\delta t_1} \\
&=& \frac{a(t_0)}{a(t_1)}.
\label{cosmoz}
\end{eqnarray}
Now one can use the redshift relation to obtain Hubble's law in the limit of small redshifts \cite{wein}.
For very small redshifts the second and higher powers of $(t-t_0)$ are ignored, and Taylor expansion about the present day value of the scale factor, $a_0$, gives
\begin{eqnarray*}
(1+z)^{-1}&\approx& (1-z) \\
&\approx& \frac{1}{a_0}[a_0 + \dot{a}|_{t_0}(t-t_0)] \\
&\approx& 1+H_0(t-t_0) \\
z &\approx& H_0(t_0-t)\\
\frac{v}{c} &\approx& H_0(t_0-t)
\end{eqnarray*}
For objects at small redshifts, proper distance from the observer at time $t_0$, given by $a(t_0)\,\chi_1$ can be approximated to $c(t_0-t)$ by using the following approximation of $\chi_1$.
\begin{eqnarray*}
\int_0^{\chi_1}{d\chi} &=& \int_{t_1}^{t_0}{\frac{c\,dt}{a(t)}} \\
 &\approx& \frac{c(t_0-t)}{a_0}
\end{eqnarray*}
Thus, $v = H_0 \, d$ in accordance with Hubble's law.

In the next chapter we take a closer look at the dynamics of an expanding universe.

\chapter{THE DYNAMIC UNIVERSE}
\label{chap3}
Einstein's equations marry the geometry of the universe to its energy-momentum content. Different forms of energy have different pressures and hence effect different changes in the scale factor of the universe with time. They also have different conservation laws. It is our aim to understand these ideas now so that we may get a better feel for what observations imply about the contents of our universe in subsequent chapters.

To begin with, we cast Einstein's modified set of equations (\ref{leineqn}) in the form of his original equations (\ref{eineqn}) so that the cosmological constant can be treated as a form of energy. To this end, we move the term with $\lambda$ in (\ref{leineqn}) to the right hand side and club it with the energy-momentum tensor. The resultant form of energy has density equal and opposite in sign to its pressure. This form of energy, which is detectable only through its gravitational effects, is called dark energy. 

There is a set of conditions, arising from physical considerations, that is imposed on the energy density and pressure of any kind of physically reasonable energy. These are called the weak, strong and dominant energy conditions. We now see if these energy conditions are obeyed by dark energy. 
The substance of the weak energy condition is the belief that physically observable energy has positive density. It translates to $T_{\mu \nu} v^{\mu}v^{\nu} \geq 0$ where $T_{\mu \nu}$ is the energy-momentum tensor in a frame of reference, $v^{\mu}$ is the 4-velocity of the observer measuring the energy density in that frame of reference.
For $T_{\nu}^{\mu}=Diagonal(\rho, -p_1, -p_2, -p_3)$, the weak energy condition is satisfied iff $\rho \geq 0$ and $\rho+p_i \geq 0$ for $i=1,2,3$. These can be obtained by considering an observer at rest in the frame of reference and an observer moving with 4-velocity such that $c\,dt/d\tau=dx^i/d\tau$ for $i=1,2,3$ respectively.
For the cosmological constant, energy density is $\lambda/8\,\pi\,G$ and pressure is $-\lambda/8\,\pi\,G$. If the cosmological constant is positive, it satisfies the weak energy condition mentioned above. Its pressure is negative, a property no known form of energy has. This has very interesting effects on the evolution of the scale factor of the universe as we will see in coming discussions.

The dominant energy condition arises from Einstein's postulate that nothing can move faster than the speed of light. This means that $J^{\mu}=T^{\mu}_{\nu} v^{\nu}$, which is the energy-momentum 4-current as seen by the observer mentioned above, should not allow for energy transfer at a speed greater than that of light. Or in other words, $J^{\mu}$ must be a future-directed timelike vector for a future-directed timelike 4-velocity $v^{\mu}$. This is formally stated as $J^{\mu}J_{\mu} \geq 0$ or $\rho \geq |p_i|$ for $i=1,2,3$. The cosmological constant can be seen to satisfy this condition as well if it is positive.

Einstein's equations can be rewritten in the form $R_{\mu \nu} = -8 \pi\,G\,(T_{\mu \nu}  - \frac{1}{2}T\,g_{\mu \nu})$ by taking the trace of (\ref{eineqn}). The strong energy condition requires that $T_{\mu \nu} v^{\mu}v^{\nu} \geq \frac{1}{2}T$. The energy condition comes from asking that $-R_{\mu \nu}v^{\mu}v^{\nu} \geq 0$ for a time-like 4-velocity $v^{\mu}$. For $T^{\mu}_{\nu}$ of the form mentioned above, it means $\rho+\Sigma_{i=1}^3{p_i} \geq 0$ and $\rho+p_i \geq 0$ for $i=1,2,3$. It must be noted that the cosmological constant violates the strong energy condition. The physical significance of this energy condition is a little harder to understand than that of the other two. For this one needs to find out what the condition $-R_{\mu \nu}v^{\mu}v^{\nu} \geq 0$ means. We start by noting that the Riemann tensor is a measure of non-commutativity of derivatives. $R_{bc}{}^a{}_du^b\,v^c\,w^d$ represents the difference in the $a^{th}$ component of vector $w$ when it is parallel transported along different directions in a path formed by vectors $u$ and $v$. Let us consider a cloud of freely falling particles, of which one particle has the velocity vector $u$. Let us consider three particles at separation vectors $v_i$ respectively where $i=1,2,3$ from this particle whose velocities are the same as that of the first particle when parallel transported through the respective $v_i$. That is, we consider particles that are in a region of uniform gravity. Then we see that $R_{bc}u^b\,u^c\,V$ is the second derivative of the volume V formed by the vectors $v_i$.\cite{baez} In the rest frame of the first particle, $R_{00}$ is a measure of the second time derivative of a unit comoving volume in a gravitational field. The said particle is at the centre of that volume. And the condition imposed on the Ricci tensor means that the second time derivative should be negative when matter density is positive, implying that the attractive nature of gravity decelerates the expansion of the comoving volume. This is analogous to a stone thrown upwards from earth slowing down due to the earth's gravitational pull. The fact that the cosmological constant does not satisfy the strong energy condition seems to indicate that it behaves differently from ordinary attractive matter. More about its unusual behaviour is discussed later in the chapter.

\section{Expansion in different scenarios}
We now proceed to study (\ref{eineqn}) applied to the Robertson-Walker metric. The two independent equations obtained are:

\begin{equation}
\left(\frac{\dot{a}}{a}\right)^2 = \frac{8 \pi G \rho}{3} - \frac{K}{a^2}
\label{hubble1}
\end{equation}

\begin{equation}
\dot{\rho}=-3 \frac{\dot{a}}{a} (\rho+p)
\label{rhodot}
\end{equation}

In order to write (\ref{hubble1}) in a different form we define certain quantities. The fraction $\frac{\dot{a}(t)}{a(t)}$ is called Hubble's constant, and written as $H(t)$. Hubble's constant at the present time is written as $H_0$. The critical density at time t is given by 
\begin{equation}
\rho_c (t) = \frac{3 H^2(t)}{8 \pi G}
\label{critdens}
\end{equation}
It is defined as the energy density of a flat universe which can yield the Hubble's constant observed at that time. The fraction of the critical density that the $i^{th}$ form of energy density forms at time $t$ is $\Omega_i^{(t)} = \rho_i/\rho_c(t)$.
$\rho_K$ is defined as $-\frac{K}{a^2}$. Using these (\ref{hubble1}) is rewritten as 

\begin{equation}
H^2(t) = H_0^2[\Sigma_i{\Omega_i^{(t)}}]
\label{hubble2}
\end{equation}

We now study the effect of different forms of energy on the evolution of the scale factor separately. To make things simpler the curvature is set to zero, $K=0$. However, it will be seen that the conclusions drawn will also be valid when curvature is not zero.
In general, one can use the equation of state, $p=w \rho$, to modify (\ref{rhodot}) to $\dot{\rho}=-3 H \rho(1+w)$ which has the solution $\rho(t) \propto a^{-3(1+w)}(t)$. This implies that $H^2(t) \propto a^{-3(1+w)}(t)$ and $a \propto t^{\frac{2}{3(1+w)}}$.

For pressureless matter, $p=0$, $w=0$ and $\rho$ falls off as the third power of the scale factor. This is expected if the total amount of matter in the universe is thought to be conserved. The scale factor grows as $t^{2/3}$.
For radiation, $p = \frac{\rho}{3}$, $w=\frac{1}{3}$ and $\rho$ falls off as the fourth power of the scale factor. The fall in energy density of radiation has a dependence on an additional factor of the scale factor owing to cosmological redshift accompanying expansion and the resultant decrease in energy of photons. The scale factor grows as $t^{1/2}$. In an expanding radiation-filled universe, the photons can be said to do work on the expanding universe and in the process lose energy, getting redshifted.
For the cosmological constant, $p=-\rho_{\lambda}$, $w=-1$. This means that $\dot{\rho_{\lambda}} = 0$ and the energy density associated with $\lambda$ is constant. It is not surprising if one notes that $\rho_{\lambda}$ is made up of constants. However, for dark energy to have a constant density as the universe expands means that dark energy is being created as the universe expands. One look at the thermodynamics of expansion can convince us that this is indeed the case. From the first law of thermodynamics, the internal energy of a system decreases by the amount of work the system has done (or equivalently increases by the amount of work done on the system by the surroundings) and increases by the product of temperature with the accompanying entropy change.
\begin{equation}
dU = -PdV + TdS
\end{equation}
For an adiabatic process, $dS=0$, there is no entropy change. The expansion of our universe can be considered a reversible isentropic process if its rate is small in comparison to the rates of the other reactions taking place within the universe. Then one notes that the sign of $dU$ is positive in an expansion if pressure is negative. Pressure of a system is usually understood as the force per unit area it exerts against its environment. The work done by a system is also understood in terms of its environment. But when the system considered is the entire universe, this description is not very useful as there is nothing outside of the universe. One could, instead, think of the positive pressure of ordinary matter and radiation as opposing their gravitational tendency to attract. Thus in an expanding universe they tend to lose energy. The negative pressure of dark energy can be understood to oppose its gravitational tendency to repel, hence, in an expanding universe the amount of dark energy increases.

As $\rho_{\lambda}$ is constant through time, so is the Hubble's constant. In this case the scale factor evolves as $e^{Ht}$. Such an empty universe with a cosmological constant is called the deSitter universe.
\begin{figure}
\centering
\includegraphics[width=150mm]{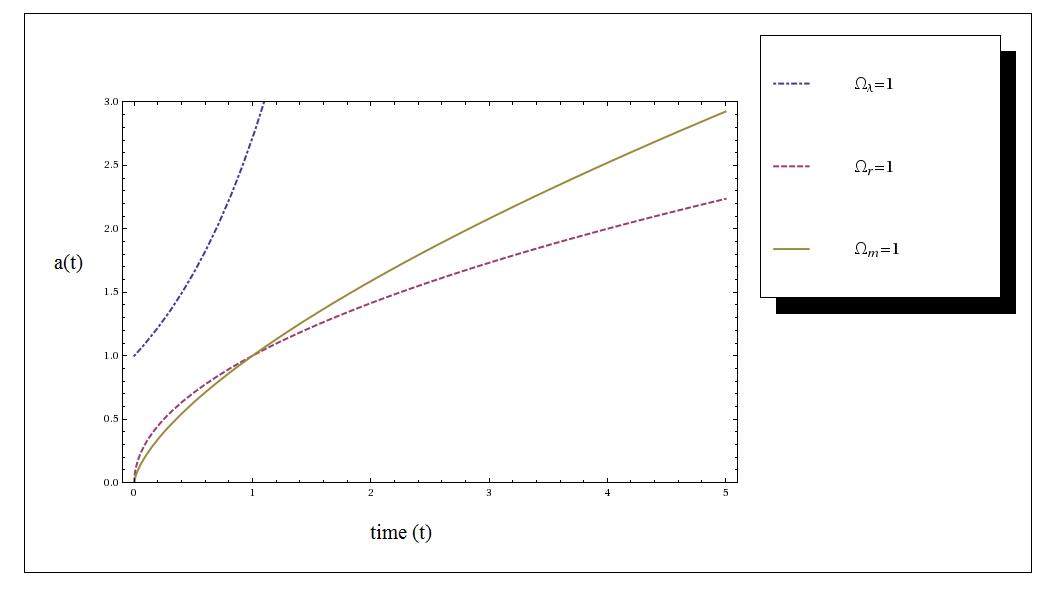}
\caption{\label{aevoln} Evolution of scale factor for different constituents of a one-component universe.}
\end{figure}
A very interesting feature of the deSitter universe can be seen from the graphs of evolution of scale factor with time. The scale factor of the deSitter universe is not zero at any time, whereas the scale factors of both a matter-filled and a radiation-filled universe have the feature of having started from a zero point. In fact, $e^{Ht}$ goes to zero only at $t=-\infty$ implying that the deSitter universe has existed forever. Thus, there is no preferred origin of time in this universe. Any time $t_0$ can be taken as the origin by the transformation $r \rightarrow e^{t_{0}}r$ and the universe appears to be the same as it is at any other time. But in the other two scenarios, there is a preferred origin of time - the time corresponding to $a=0$. One cannot talk sensibly of a time before this because the very existence of the universe became a reality only at that origin. Such a birth of the universe from a single point is termed the Big Bang. At the singularity of $a=0$ density of the universe, the terms of the Ricci tensor and the Ricci scalar all diverge. The general theory of relativity cannot be said to make sense at the singularity. But this unseemly breakdown of our theories is avoided in a universe with a positive cosmological constant.
We note that $H^2 = \lambda$ allows for $a=\cosh{\sqrt{\lambda}\,t}$ as a solution. $\cosh{t}$ has a minimum at $t=0$. What this implies for the universe is that before $t=0$ the universe contracts and post $t=0$ the universe expands but a singularity is avoided at time $t=0$. This scenario is sometimes called the Big Bounce in contrast to the Big Bang.

To make another interesting point about the cosmological constant, we rewrite (\ref{rhodot}) by substituting for $\rho$ from (\ref{hubble1}) as follows:
\begin{equation}
3\, \frac{\ddot{a}}{a} = - 4 \,\pi\, G \,(\rho + 3\,p)
\label{acce}
\end{equation}
 
\begin{figure}
\centering
\includegraphics[width=150mm]{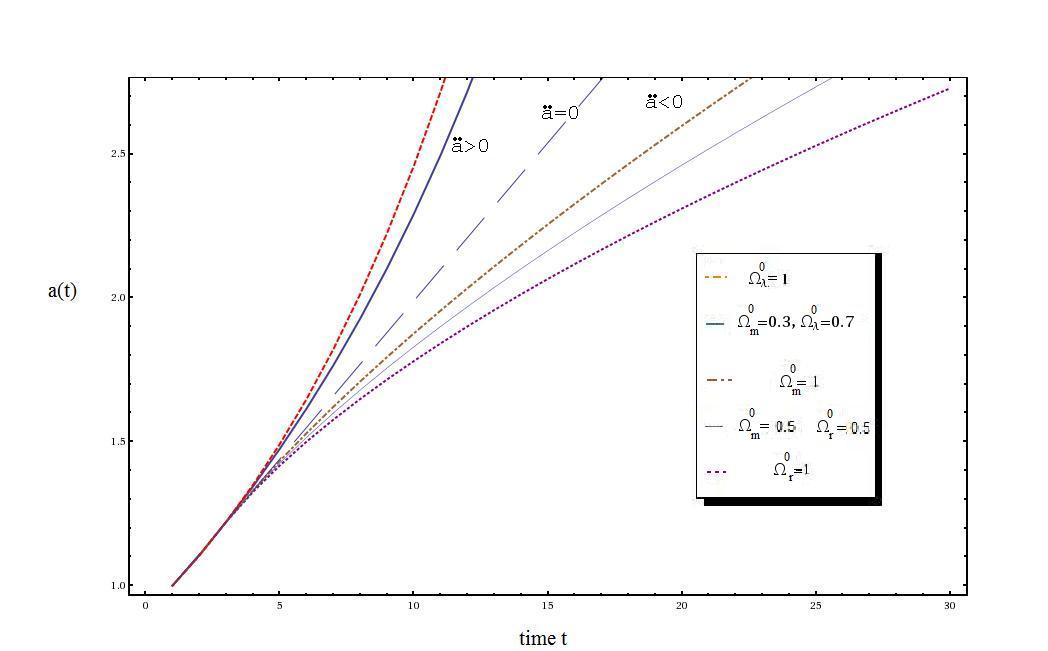}
\caption{\label{aevolnmix} Evolution of scale factor with time in different scenarios.}
\end{figure}

One immediately sees that both matter and radiation predict a negative value for $\ddot{a}$ whereas the cosmological constant predicts a positive value for the same. Just as a ball thrown upwards from earth is slowed down due to the attractive pull of earth, the expansion of the universe is slowed down by the attractive gravitational effect of ordinary matter and radiation. But it seems as though dark energy actually repels as it accelerates expansion. We had a hint of this effect of dark energy when we discussed its violation of the strong energy condition. It is noted that only energy forms with negative enough pressure $(w < -\frac{1}{3})$ can accelerate the expansion of the universe. One can also make out from figure \ref{aevolnmix} that evolution curves of models with a non-zero positive cosmological constant fall on the opposite side of the straight line curve (that has no acceleration) to that containing evolution curves of models with no dark energy.

It is also clear that for $\ddot{a}$ to be zero, as in a static universe, one would require the existence of some form of energy with negative pressure, $\rho = -3\,p$, or $w=-\frac{1}{3}$. Since this condition was not fulfilled by any known form of energy then, Einstein introduced the cosmological constant into his equations. An important feature of Einstein's static universe can be deduced by studying (\ref{acce}) - its instability.\cite{Eddington} In Einstein's static spherical universe, the values of the cosmological constant, $\lambda$, and density of pressureless matter, $\rho$, are such that their effects exactly cancel out. But such a balance is, in fact, very delicate making such a universe unstable. If $\lambda$ is slightly larger, then the universe gets a positive acceleration and expands, decreasing the density in the process and hence, increasing the acceleration further. Similarly, if $\lambda$ is smaller, contraction of the universe causes the density to increase and further fuel the deceleration. Thus, Einstein's static spherical universe is unstable.

It is interesting to see what an empty universe devoid of any form of energy will look like. For zero curvature, ($\ref{hubble1}$) and ($\ref{acce}$) yield constant scale factor as a solution. But for $K=-1$, the equations actually yield a non-static solution of $a \propto t$. This universe is called Milne universe. What is interesting about it is that with the coordinate transformations $R=t\,sinh\chi$ and $T=t\,cosh\chi$, the line element of the Milne universe looks like the following:
\begin{equation}
d\tau^2=c^2\,dT^2-[dR^2+R^2\,(d\theta^2+sin^2\theta\,d\phi^2)]
\end{equation}
This is just a quadrant of the Minkowski spacetime where both $T$ and $R$ are positive.

We have looked at solutions of Einstein's equations assuming the universe to be filled with only one type of energy. This was done to understand the effect of each individually. When there is a mixture of various forms of energy, the evolution of the scale factor would depend on the fraction each of these forms of the total density. What is interesting is that, due to the difference in the way the densities of various energy forms evolve, the fraction each forms of the total energy content of the universe also evolves through time. For arguments sake, let us assume that the world today has a greater density of matter than the energy density of radiation. If one goes backward in time, the energy density of radiation rises faster than that of matter. So, there could be some point in time in the past when radiation energy density was the same as matter density. The following equation helps us determine the redshift to that time if the current ratio of densities is known.
\begin{eqnarray}
\rho_r(t) &=& \rho_m(t) \\
\rho_r^0 \frac{a_0^4}{a^4(t)} &=& \rho_m^0 \frac{a_0^3}{a^3(t)} \\
\frac{\rho_m^0}{\rho_r^0} &=& \frac{a_0}{a(t)} \\
	&=& 1+z
\label{radndom}
\end{eqnarray}
At higher redshifts radiation energy would have been the dominant form of energy. 
Similarly the start of the period of dark energy domination over the matter density can be calculated by $\rho_m^0 \, (1+z)^3 = \rho_{\lambda}$. From current estimates of the relative fractions of dark energy density, radiation density and matter density, which will be given later, one can estimate that the universe was radiation-dominated till a redshift of about $3000$, matter-dominated till around $z=1/3$ and dark energy-dominated ever since. So, many calculations can be simplified by taking into account only the dominant forms of energy and ignoring the others. It should be noted that this part of our study does not require the assumption of a flat universe.

One can also calculate when accelerated expansion of the universe began from (\ref{acce}).
\begin{eqnarray}
\rho_m(t) + \rho_{\lambda}(t) - 3 p_{\lambda}(t) &\leq& 0 \\
\rho_m(t) &\leq& 2 \rho_{\lambda}(t) \\
\rho_m^0\,(1+z)^3 &\leq& \rho_{\lambda}^0 \\
z &\leq& \left(\frac{\rho_{\lambda}^0}{\rho_m^0}\right)^{1/3}-1
\end{eqnarray}
If the estimates of current densities of dark energy and matter are correct, this implies that accelerated expansion of the universe began at a redshift of about $0.67$, that is even before the universe switched over from being matter-dominated to dark energy-dominated.

Before we move on, it is useful to compare the evolution of the scale factor in various models of the universe with the evolution of the position of our stone in the introductory chapter.
The motion of the stone with zero acceleration and constant velocity in the absence of earth, finds an analogue in the Milne universe. The scale factor of a matter-filled universe evolves just the way the radial coordinate of the stone with zero total energy evolves, as $t^{2/3}$. But models with positive acceleration of $a$ find no analogy in our example of the motion of the stone.

We now have sufficient background to examine certain important astronomical observations.

\chapter{THE ACCELERATING UNIVERSE}
\label{chap4}
Edwin Hubble's data was the first observational evidence for an expanding universe. His data consisted chiefly of luminous objects within a few hundred Megaparsecs (Mpc) of earth, possessing velocities of around 1000 km/s. 
The original plot \cite{edhub} had a large deviation from the best fit line. There was a need to collect more data and probe deeper into space to see if the relation between velocity and distance held. With improvements in observation techniques and newer developments in measuring astronomical distances this was made possible. In astronomy, since the distance scales are huge, one needs to make use of indirect methods of measuring distances. From common experience, one knows that a candle, when moved away from an observer, appears dimmer with increasing distance. Thus the apparent luminosity of a luminous object can be a measure of its distance from the observer if its absolute luminosity is known. In other words, the relation between distance, brightness perceived by the observer and actual intrinsic brightness can be determined precisely if the geometry of space is known. Similarly, any object appears to become smaller as it is placed further away from the observer. Thus, brightness and angular size of luminous objects perceived by observers are used to estimate distances to those objects. The study of red-shifts of supernovae to estimate distances makes use of the former idea, while the study of anisotropies in the cosmic background radiation makes use of the latter. We understand the supernova data in this chapter and the cosmic background radiation data in the next.

\section{Standard candles}
In order to accurately estimate distances by observing luminous objects, the intrinsic luminosity of these objects must be known. A luminous object which can be uniquely identified by its characteristic features and has uniform absolute luminosity across its samples is called a standard candle. The brightest standard candles in use today for distance estimation are type \Rmnum{1}a supernovae where \Rmnum{1}a specifies the spectral class of the supernova. This specific type of supernovae has been observed to have very high absolute peak luminosity \footnote{Peak luminosity of a supernova refers to the value of luminosity when it is brightest.}, about a few billion times brighter than the sun. In addition, the peak luminosities are fairly uniform over many such supernovae. This uniformity arises from the process that leads to type \Rmnum{1}a supernova explosions. A type \Rmnum{1} supernova usually occurs when a small white dwarf star uniformly accreting matter from some nearby source, exceeds a certain critical mass, at which point the outward electron degeneracy pressure of the gas in the star is no longer sufficient to counter the effect of gravity. The white dwarf then begins to collapse, increasing the temperature in its core. This results in uncontrolled nuclear fusion, causing an explosion which is the supernova. Since the masses of stars that explode in this fashion are very close to the critical mass, and hence nearly uniform, their absolute luminosities also are. Hence, the magnitude and uniformity of their intrinsic brightness make type Ia supernovae good standard candles. Supernovae of type \Rmnum{2} are not good standard candles as they differ widely in their absolute luminosities and are intrinsically dimmer than those of type \Rmnum{1}a. They occur when massive stars have cores heavier than the Chandrasekhar limit after their hydrogen fusion stage. But the processes leading up to the explosion are not yet clearly understood. Other potential bright sources of radiation that do not make good standard candles are active galaxies since they cannot be assigned standard luminosities. Their luminosities are known to evolve with time.

\subsection{Calculation of Chandrasekhar's limit}
The critical mass mentioned above is called Chandrasekhar's limit and it can be computed as follows.
A star, while fusing hydrogen to form helium in its core, maintains equilibrium between the opposing effects of gravity and outward gas and radiation pressure. Once the star runs out of hydrogen in the core and the fusion process stops, gravity wins the tug of war and the star begins to collapse in on itself. Does this collapse proceed indefinitely? The answer depends on the amount of matter that is collapsing. As the star - more correctly, its core - collapses, the density of matter increases and the mean free path of the constituents decreases. At the temperature in the star core, which is of the order of ten million kelvin (or of the order of keV), constituent matter is ionised. When the density is sufficiently high, the mean free path becomes as small as the de-Broglie wavelength of an electron. Two electrons at such a separation cannot be resolved. However, Pauli's exclusion principle bars two electrons of the same spin from occupying the same quantum state simultaneously. The electrons are, thus, forced into higher energy states increasing gas pressure. They are then said to form a degenerate gas. Another way to understand degeneracy pressure is by noting that from the uncertainty principle, one knows that the more precisely a particle's position is known, the less precise is the knowledge of its momentum. Thus, at very high densities, particles must have large momenta uncertainties contributing to the pressure of the gas that resists compression. 

The highest energy level occupied by an electron in a degenerate gas at 0 K temperature is called the Fermi energy. This can be easily calculated from the number of electrons in the system per unit volume $n_e = 2 \times 4\, \pi\, p_0^3/8 \,\hbar^3$.

Degeneracy pressure of electron gas can be calculated in the following manner.
$P=\frac{dp/dt}{A}$
The rate of change in momentum per unit area of a surface is given by
\begin{equation}
\int_0^{\infty}{2\,p\,cos\theta\, n(p,\phi,\theta)\, v\, sin\theta\, d\theta\, d\phi\, dv}
\end{equation}
where $n(v,\phi,\theta)dv$ is the number of particles per unit volume coming from the $(\theta,\phi)$ direction with velocity in the interval v to $v+dv$ in the solid angle $d\Omega=sin\theta \,d\theta\, d\phi$.
The change in momentum of a particle bombarding the surface with momentum p and at angle $\theta$ is $2 \times p\, cos\theta$ assuming it undergoes a mirror reflection.
The integral can be simplified to $\frac{1}{3}\int{n(v)vdv}$ if one assumes that the distribution of particles hitting the surface is isotropic and hence is a function only of speed.
For a degenerate gas, $n(p)dp=2 \times 4\, \pi\, p^2\, dp/\hbar^3$, which has been obtained using both Pauli's exclusion principle and Heisenberg's uncertainty principle. $n(v)dv$ can equivalently be written as $n(p)dp$ where p is the momentum corresponding to speed v.
The pressure of a degenerate gas is thus 
\begin{equation}
\frac{2 \times 4\, \pi}{3\hbar^3}\int_0^{p_0}{v\, p^2\, dp}
\end{equation}
At low velocities $v \approx p/m$, $P \propto p_0^5 \propto n_e^{5/3} \propto \frac{M^{5/3}}{R^5}$.\\
But as velocities become relativistic $v \approx c$, $P \propto p_0^4 \propto n_e^{4/3} \propto \frac{M^{4/3}}{R^4}$.\\
The gravitational pressure at the core due to mass is $P_g \propto \frac{M^2}{R^4}$.\\
Usually, in a cool white dwarf star that has just begun contracting after exhausting hydrogen in its core, the low velocity approximation works fine. But as it contracts further and heats up, relativistic effects need to be taken into account. If the mass is small enough, the degeneracy pressure stops further collapse. But if the mass is greater than a critical mass, the star continues to shrink in size, and both the degeneracy and gravitational pressure increase. The critical mass can be obtained by equating the degeneracy pressure of a relativistic configuration and its gravitational pressure and the value obtained is around $1.4$ solar masses. This is called the Chandrasekhar limit after the astrophysicist who originally calculated its value.

\section{The accelerating universe}
Once the apparent luminosities of many different type \Rmnum{1}a supernovae have been obtained, it is easy to determine which cosmological model best explains the data, as one can derive a relation between the flux of radiation from a supernova received at earth and the amount by which light has been redshifted. We start by defining luminosity distance $d_l$ by extending the relation in Euclidean space between distance to an object, its luminosity and flux at the observer, to any arbitrary spacetime.
\begin{equation}
d_l^2 = \frac{L_s}{4 \pi F}
\label{lumdis}
\end{equation}
In an expanding universe, two effects of the expansion on the flux at the observer must be considered. As universe has expanded in the interval between emission of photon at source and its detection by the observer, its wavelength has increased and the frequency of radiation has decreased. Thus, not only does the amount of energy reaching the observer lessen, so does the rate. Let O be the surface of all points equidistant from the source, and containing the coordinate point of the observer. Then, the rate of energy received at surface O is related to luminosity at source in the following manner, assuming of course, that no energy is absorbed and lost between emission and detection.
\begin{eqnarray}
\frac{L_s}{L_0} & = & \frac{\Delta E_s/\Delta t_s}{\Delta E_0\Delta t_0} \\
\nonumber	& = & \frac{\nu_s}{\nu_0}^2 \\
\nonumber	& = & (1 + z)^2
\end{eqnarray}

From the Robertson-Walker metric, the flux at the observer is $F = L_0/4\,\pi\, a_0\, f_K^2(\chi)$, where $\chi$ is the radial coordinate of the source. Using this, one obtains the luminosity distance as \\ $d_l = a_0 \,f_K(\chi)\, (1 + z)$.
It is useful to write quantities in terms of redshifts because redshifts can be very precisely measured. The radial coordinate $\chi$ can be written as a function of redshift by noting that photons follow null geodesics. 
$d\tau^2 = dt^2 - a^2(t)d\chi^2 = 0$.
\begin{equation}
\int_0^\chi{d\chi} = \int_{t_1}^{t_0}{\frac{dt}{a(t)}} 
\end{equation}

\begin{eqnarray*}
1+z     &=& \frac{a(t_0)}{a(t)} \\
\dot{z} &=& \frac{-a(t_0)}{a(t)}\frac{\dot{a}(t)}{a(t)} \\
        &=& -(1+z)H(z) \\
    dt  &=& \frac{-dz}{(1+z)H(z)}
\end{eqnarray*}

\begin{equation}
\chi = \int_0^z{\frac{dz'}{a_0 H(z')}} 
\label{chi(z)}
\end{equation}

\begin{figure}
\includegraphics[width=150mm]{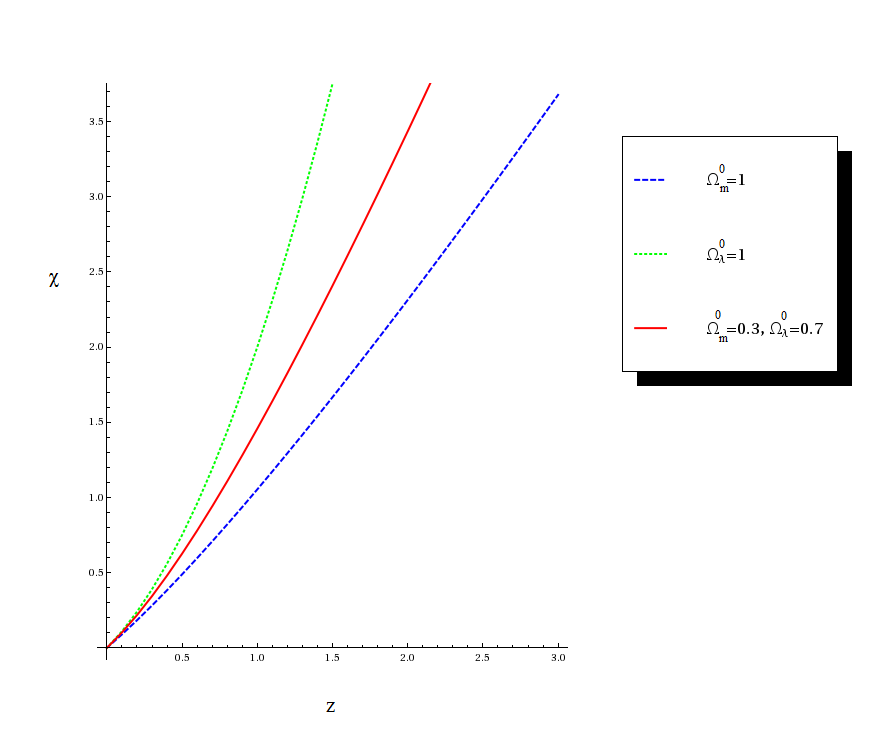}
\caption{Coordinate distance as a function of redshift in a flat universe.}
\label{chiz}
\end{figure}
In terms of the densities of various constituents of the universe the Hubble's constant is written as follows:
\begin{equation}
H(z) = H_0 \sqrt{\Sigma_n \Omega_n^{(0)} (1+z)^{3(1+w_n)}}
\label{hubble3}
\end{equation}
Thus, we see that the value of $\chi$ depends on the energy density composition of the universe. For the moment we assume the universe is flat with $\Omega_K = 0$. The proof for the flatness of the universe comes from anisotropies in the cosmic background radiation, which we shall learn about in the following chapter. The effect of a non-zero, positive cosmological constant on the radial coordinate $\chi$ can be understood as follows. The acceleration of the universe increases with increase in $\lambda$. At a given redshift, Hubble's constant is smaller for a universe with a non-zero, positive cosmological constant than for one without. This has the effect of increasing the integrand in (\ref{chi(z)}). In effect, a luminous object appears to be further away and hence, dimmer in a universe with positive $\lambda$ than in one without. And this is precisely the effect that was observed by groups studying supernovae at large distances, the High-z Supernova Search Team and the Supernova Cosmology Project.\cite{supernova} The cosmological model that best fits the supernova data has dark energy forming about $70 \%$ of the critical density of the universe today and matter, both baryonic and cold, forming the remaining $30 \%$ with negligible contribution from radiation.\cite{edcope}
\begin{figure}
\includegraphics[width=150mm]{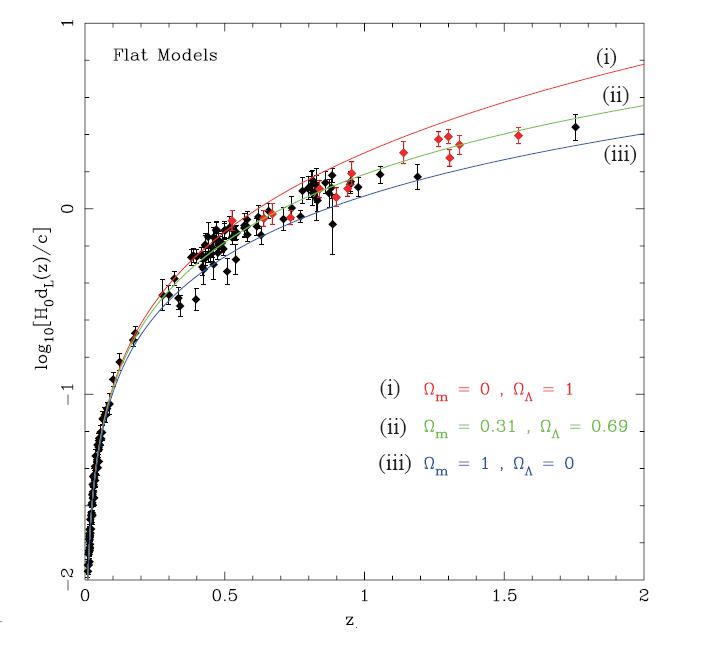}
\caption{The log of luminosity distance versus redshift data for supernovae obtained by the Supernova Search Team Collaboration and the Hubble Space Telescope, taken from \cite{edcope}.}
\label{superdata}
\end{figure}

To see more lucidly the effect of the acceleration of the universe on the luminosity distance, one may obtain the luminosity distance as a series in powers of redshift.
\begin{eqnarray*}
(1+z)^{-1} &=& \frac{a(t)}{a_0} \\
1-z+z^2... &=& 1+H_0\,(t-t_0)+\frac{\ddot{a}}{a}|_{t_0} \frac{(t-t_0)^2}{2!} ... \\
z &=& H_0\,(t_0-t)+H_0^2\,(t_0-t)^2\,[1-\frac{\ddot{a}}{a}|_{t_0}/(2\,H_0^2)] ... \\
H_0\,(t_0-t) &=& z-z^2\,[1-\frac{\ddot{a}}{a}|_{t_0}/(2\,H_0^2)] ...
\end{eqnarray*}
We also obtain $\chi$ as a power series.
\begin{eqnarray*}
\chi &=& \int_{t_1}^{t_0}{\frac{dt}{a(t)}} \\
 &=& \int_{t_1}^{t_0}{\frac{dt}{a_0\,(1+H_0(t-t_0)...)}} \\
 &=& \frac{1}{a_0}\int_{t_1}^{t_0}{dt(1-H_0(t-t_0)...)} \\
 &=& H_0^{-1}/a_0\,[ H_0\,(t_0-t_1)+ \frac{H_0^2\,(t_0-t_1)^2}{2} ...]
\end{eqnarray*}
Using both the above expansions, one can write the luminosity distance retaining only terms up to second order in $z$ as follows:
\begin{eqnarray*}
d_l &=& a_0\,f_K(\chi)\,(1+z) \\
    &\sim& (1+z)\,H_0^{-1}\,[ H_0\,(t_0-t_1)+ \frac{H_0^2\,(t_0-t_1)^2}{2}...] \\
    &\sim& H_0^{-1}\,[ z+\frac{z^2}{2}(1+\frac{\ddot{a}}{a}|_{t_0}/(H_0^2))...]
\end{eqnarray*}
From the above relation, it is very easy to see that for a given redshift, a positive acceleration produces a greater luminosity distance, thereby decreasing the flux from the source reaching the observer. This makes the luminous object appear dimmer.

Yet another way to understand the supernova data is in terms of how much the universe has aged since the supernova exploded. From figure \ref{age} that shows the age of the universe for various models, one sees that a given redshift corresponds to an older time in the past for an accelerating universe than for the other cases. Since the speed of light is constant, light has traveled a greater distance in an accelerating universe. Thus, for the same redshift, a supernova is more distant in an accelerating universe and hence, dimmer, than in other non-accelerating universe models.

\chapter{GEOMETRY OF THE UNIVERSE}
\label{chap5}

In this chapter we look at the problem of determining the global curvature of the universe. As has been mentioned in Chapter 3, the energy density of the universe decides the curvature of the universe. Let us call the ratio of the total energy density of the universe to the critical energy density of the universe at a particular time $\Omega(t)$. According to (\ref{hubble1}), for $\Omega < 1$, $K < 0$, for $\Omega > 1$, $K > 0$ and for $\Omega = 1$, $K = 0$. One way, then, of determining K would be to calculate observed densities of different forms of energy and add them up to see which of the above conditions they satisfy for the present-day value of Hubble's constant.

Another way of determining curvature is to look for geometric relationships that are affected by the value of K. For instance, one knows that the sum of three angles of a triangle is $180^{\,\circ}$ only on flat space. On a positively curved surface like that of a sphere it is greater than $180^{\,\circ}$ while on a negatively curved surface it is less than $180^{\,\circ}$. One such relationship that astronomers study is the one between distance to an object and the angle it subtends across the line of sight. In flat space, one knows that an object with linear extent $x$ and at distance $r$ from the observer subtends an angle $x/r$ in radians at the observer. However, in curved spaces this relationship does not hold good, at least not to second and higher order approximations. In cosmology, we are interested in how the Robertson-Walker metric, (\ref{rwmetric}), relates the mentioned quantities.
At time $t$, $x=a(t)f_K(\chi)\theta$ is the linear extent of an object at coordinate $\chi$ subtending angle $\theta$ at the origin of the coordinate system. Thus, if one could find a standard length and plot the angle it subtends at an observer at the origin versus its coordinate for different values of $\chi$, the resultant curve will indicate the curvature of the universe. Ideally one would want such a standard length, called standard ruler, at many different distances from the observer to determine curvature of the universe. A standard ruler must have a large linear extent in order to have appreciable angular scale at large distances and it must not evolve with time except through its dependence on the scale factor. That is, a length $x$ at $(t, \chi)$ must appear as length $x\, a(t_1)/a(t)$ at $(t_1, \chi)$. It is also necessary to either have some distance independent method of estimating that linear extent, or be able to measure the standard ruler at a distance close enough to the observer so that curvature does not affect its extent much. (The latter method was adopted in finding absolute luminosities of the standard candles, type \Rmnum{1}a supernovae.) These requirements have made standard rulers rare. Since large objects like galaxies are known to evolve with time, they make bad standard rulers. The most successful standard ruler has been the correlation function in the anisotropy of the cosmic background radiation. 

\section{Discovery of Cosmic Microwave Background Radiation}
The cosmic microwave background radiation (henceforth referred to as CMBR) was discovered in 1964 by radio astronomers Arno Penzias and Robert Wilson \cite{penwil} quite by accident. Despite their best efforts to reduce noise in their radio wave detector, they were left with a background noise. The noise was uniform through the day and night, and did not change when the direction of the antenna was changed. They could not zero in on any source of radiation in the sky either, or rather, the whole sky seemed to radiate. This led to the conclusion that the noise had a cosmic origin and was actually radiation from a very distant past. The cosmological significance of this noise was explained by Dicke, Peebles, Roll and Wilkinson \cite{peebles}, who had, at the time of the discovery, been trying to detect this radiation. Since the discovery in 1964, it has been confirmed through many observations that the background is, in fact, an almost perfect blackbody curve at an approximate temperature of 2.7 Kelvin.\cite{wein3mins} \footnote{There had been earlier indications of presence of cosmic radiation. But before 1964, the link between these observations and their cosmic significance had not been ascertained. Subsequent to the discovery by Penzias and Wilson, it was recognised that the CMBR could account for these observations as well. As Arthur Kosowsky notes in \cite{earlycmb} ``Prior to this, Andrew McKellar (1940) had observed the population of excited rotational states of CN molecules in interstellar absorption
lines, concluding that it was consistent with being in thermal equilibrium with a temperature
of around 2.3 Kelvin. Walter Adams also made similar measurements (1941). Its significance
was unappreciated and the result essentially forgotten, possibly because World War II had
begun to divert much of the world
cm corresponding to a blackbody temperature of 4 $\pm$ 3 K independent of direction. The significance of this measurement was not realized,
amazingly, until 1983!"}

To understand the origin of the radiation one needs to look at the history of the universe. Assuming the universe is currently expanding, the extrapolation backwards in time would mean that the universe was denser and hotter in the past. At some point in time density of matter and temperature must have been high enough for protons to overcome their Coulomb repulsion, facilitating formation of nuclei, the same process that happens in stellar cores today. If this was the case, why do we not observe large abundances of all kinds of elements in our universe today? Why is it that $75 \%$ of the observable universe today is made up of hydrogen? \footnote{The spectra of stars and interstellar matter suggest that most observable matter in the universe consists of hydrogen and helium nuclei and higher elements are found only in trace amounts. It is now believed that these higher elements are formed in the explosive last stages in the lifecycle of very massive stars.} This means that there is a need to explain why heavier elements did not form in the very early stages of evolution of universe when temperature and density of matter were high enough to allow such formation. The reason is the presence of a very large number of highly energetic photons which constantly collided with nuclei breaking them apart as soon as they were formed, thus keeping the number of heavy nuclei low. 
At such high temperatures electrons, nuclei and photons constantly interacted with each other and were in equilibrium. By equilibrium, one means that the rates of interactions of these constituents of the universe was greater than the rate of expansion of the universe. As the universe cooled, reaction rates dropped and constituents of the plasma dropped out of equilibrium. We are interested in the period just prior to when matter and photons stopped interacting and went out of equilibrium. This falling out of equilibrium was facilitated by the universe cooling down enough to form neutral hydrogen atoms, that is, when the temperature dropped to below an equivalent of around $13.2$ eV or $3 \times 10^3$ K. This period is known as the surface of last scatter since Compton and Coulomb scattering were the most prominent interactions taking place before matter and photons suddenly dropped out of equilibrium.  Thus, photons were no longer impeded by free electrons and the universe became transparent to electromagnetic radiation. It is the remnant of this isotropic radiation that is widely believed to be the source of noise in Penzias' and Wilson's antenna. The spectrum of cosmic background radiation being studied today thus bears the signature of events at the surface of last scatter.\cite{earlyuniv}

\section{Redshifted blackbody spectrum}
At this juncture, it is useful to see how the freely moving photons have been affected since last scatter. Let us see how the spectrum of radiation has changed in the meantime.
For blackbody radiation at temperature T, the number density of photons with frequency in the interval $\nu$ and $\nu + d\nu$ is given by 
\begin{equation}
n_T(\nu)d\nu = \frac{8 \pi \nu^2 d\nu}{e^{h\nu/kT}-1}.
\end{equation}
In an expanding universe, photons suffer cosmological redshift. We assume that this is the only effect on the photons, and that their numbers are not changed by other processes. Photons with frequency $\nu$ today must have had frequency $\nu(1+z)$ at redshift z. Also, since size of the universe increases, but not the number of photons, density decreases by a factor of $(1+z)^3$.

\begin{eqnarray}
n_{T_0}(\nu)d\nu &=& \frac{n_T(\nu(1+z))d\nu(1+z)}{(1+z)^3} \\
		 &=& \frac{8 \pi {\nu(1+z)}^2 d\nu(1+z)}{[e^{\frac{h\nu(1+z)}{kT}}-1](1+z)^3} \\
		 &=& \frac{8 \pi \nu^2 d\nu}{e^{\frac{h\nu}{kT/(1+z)}}-1}
\end{eqnarray}
Thus, we find that the blackbody spectrum retains its shape with a lowered temperature of $T/(1+z)$. Since, CMBR is almost perfect blackbody radiation at $2.7 K$, one concludes that the radiation that left the surface of last scatter also had a blackbody spectrum, but of temperature around $3 \times 10^3 K$. This gives the surface of last scatter a redshift of approximately $1100$.

\section{Anisotropies in CMBR}
The radiation discovered by Penzias and Wilson is almost isotropic. But better instrumentation allowed for finer angular resolution and soon anisotropies were detected in the spectrum.\cite{anisotropies}

\begin{figure}
\centering
\includegraphics[width=150mm]{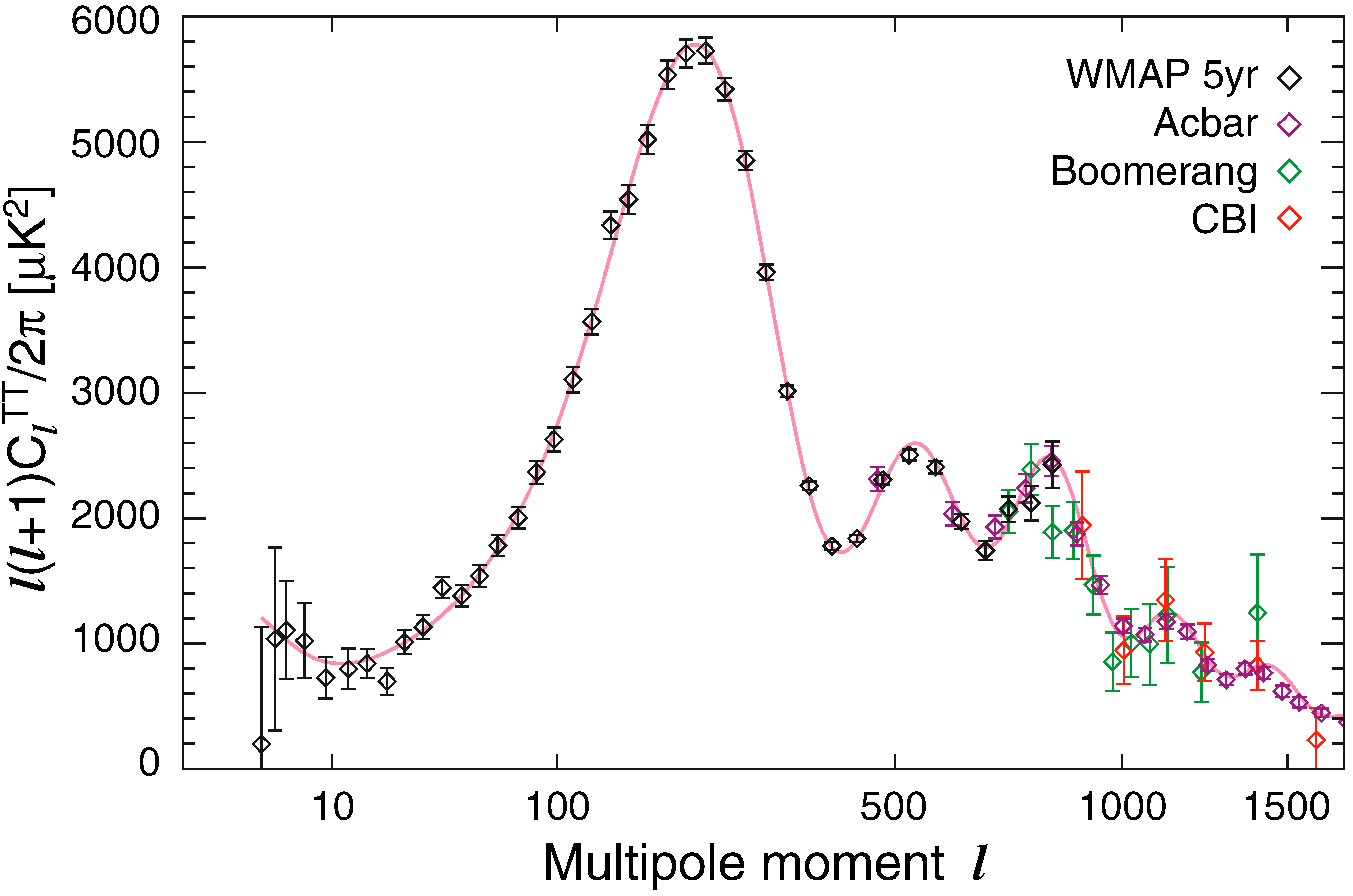}
\caption{Anisotropies in the CMBR spectrum obtained by the WMAP Science team.\cite{wmapscteam}}
\label{cmbaniso}
\end{figure}

There are many different causes for anisotropy in the CMBR spectrum.\cite{weincosmo}
The simplest is dipole anisotropy that arises due to the motion of earth. This causes the earth to receive greater flux of photons in the direction of motion than in the opposite direction. If CMBR is considered to be perfectly isotropic and homogeneous, this effect can be used to determine the velocity of motion of earth through that isotropic and homogeneous background.
The Sunyaev-Zel'dovich effect is another cause of anisotropy of the CMBR, and is due to the scattering of this radiation by electrons in interstellar matter along the line of sight. The anisotropy thus produced depends on the frequency of light in a certain manner, and hence, can be distinguished from other types of anisotropies.

The above two effects are secondary effects in the sense that they have caused anisotropy in the CMBR spectrum in the recent past of the universe. These aid us in understanding our immediate neighbourhood in the universe. Those anisotropies that are a signature of the universe at times close to the time of last scattering are called primary anisotropies. These give very useful information, as we shall see, about the curvature and constituents of the universe.
Primary anisotropies could be caused by variations in the early universe plasma density or by Doppler shift due to motion of photons in that plasma. Those caused by gravitational potentials in the plasma are included in the Sachs-Wolfe effect. The integrated Sachs-Wolfe effect is also an effect of the gravitational field on photons, but the potentials referred to here are the time dependent ones that the photon has encountered during its journey since last scatter. The time dependent nature of the potential wells ensures that the redshift of a photon as it climbs out of a potential well does not exactly negate the effect of the blueshift it suffers as it falls into the well. The nature of the potential wells depends on the expansion of the universe, and hence, the fractions of different energy densities. It also depends on reactions taking place within the universe which dictate the abundances of various species at a certain stage in the evolution of the universe.

Figure \ref{cmbaniso} shows the plot of anisotropies observed in the CMBR. The plot is a correlation of temperature fluctuations at two different angles for various values of their angular separation. The x-axis is a measure of the reciprocal of angular separation that is, it is a measure of angular frequency. Studying the deviation in temperature from the mean might not yield any interesting result since $<\Delta T> = 0$ ($<A>$ refers to the average value of a quantity A). That is, the sum of deviations from a mean on either side of the mean cancel out.  However, things could get interesting when $<\Delta T(\theta) \Delta T(\theta + \alpha)>$ is calculated for various values of $\alpha$ and this is just what was observed in the CMBR spectrum. It appears that there is some periodicity in the pattern of deviation of the temperature from the mean as one scans the sky. The anisotropies look like echoes of some wave-like disturbance from the past. In order to uncover what these patterns say about the universe, we need to understand better what caused them.

\section{Jeans criterion}
A gas in a box has two opposing tendencies; to expand owing to its outward pressure generated by kinetic energy of the molecules and to contract owing to the gravitational attraction of the molecules. Which tendency wins depends on the temperature and mass of the gas contained in the box.

Consider a static, homogeneous, isotropic universe that is allowed by a positive cosmological constant with uniform matter density $\rho_0$. Imagine that density fluctuations are introduced into this. Places with high density tend to attract surrounding matter, thus making the rare surroundings even rarer. However, if the gas is hot enough, it will have enough pressure to withstand the tendency to collapse. In other words, the gas pressure can prevent the density fluctuation from growing more pronounced. However, the larger the size of an overdense region, the greater is the gravitational pull towards collapse. Thus for a given temperature and density of matter, there is a size beyond which collapse is inevitable. This is called Jeans length. Below this cutoff, the characteristic time of collapse of the region is greater than the time required for sound to travel through the region. Thus, if a disturbance is set up with wavelength smaller than the Jeans length, the oscillations survive. For larger wavelengths, there will be no oscillation as pressure, which acts as the restoring force, is insufficient to prevent collapse.
The Jeans length is the wavelength below which stable oscillations occur and gravitational collapse is prevented.

For a cloud of ideal gas at a given temperature and density, the condition for hydrostatic equilibrium, so that it has not yet succumbed to its own gravity and collapsed, is 
\begin{eqnarray*}
Gravitational\, pressure &=& \int_0^R{\frac{G\, \rho\, 4 \pi\, r^3 \rho\, 4 \pi\, r^2 dr }{3 \times 4\, \pi\, r^2\, r^2}} \\
&=& \frac{2 \pi G \rho^2 R^2}{3} \\
&=& \frac{\rho k T}{m} \\
&=& Gas\, pressure
\end{eqnarray*}
This gives $R = \sqrt{3\, k\, T/2\,\pi\,\rho\,G\,m}$ where m is the mean molecular weight of the constituents of the ideal gas.

The early universe was a dense plasma of ionised matter and radiation. The constituents of the plasma experience the same opposing tendencies that the gas in the box experience. If the conditions of the universe are known at that time, one could estimate the largest wavelength oscillation that can be supported in the medium. But how do we know for sure what constituted the plasma at the time of recombination? The answer lies in the fact that the conditions in the universe then are responsible for the conditions in the universe now. By this we mean that whatever we assume about the constituents and conditions of the early universe must be able to evolve into what we see today. The key to answering our question about conditions in the universe at the time of last scattering is the process of nucleosynthesis.

\section{Nucleosynthesis}
The idea that elements were formed in the highly dense and hot environment in the universe right after the Big Bang was first put forward by Ralph Alpher and George Gamow. \footnote{Hans Bethe's name was added as an author to their paper to give it an alphabetic ring. Bethe went on to do seminal work on nucleosynthesis in stars and correctly postulated that nuclear fusion was the source of stellar energy.}
The success of nucleosynthesis lies in the fact that it is able to account for the current observation that $75\%$ of baryonic matter by weight is elemental or ionic hydrogen, $25\%$ is helium and there are very small traces of higher elements. 

Without getting into the details of nucleosynthesis, we look at some of its key features.\cite{earlyuniv}
\begin{itemize}
\item The main idea behind nucleosynthesis is that knowing the conditions of the universe and cross-sections of relevant reactions allows one to predict elemental abundances. 
\item Nucleosynthesis begins when the universe is cool enough to contain protons and neutrons. The sequence of events in early universe nucleosynthesis depends crucially on the photon-to-baryon number ratio and the proton-to-neutron number ratio at the start of the process, and the rate at which temperature drops owing to the expansion of the universe. Once the initial conditions are described, it is possible to predict what happens if one knows how the universe evolves, since the reactions between the constituents of the universe at that time, neutrons, photons and protons, are well understood.
\item  The second lightest stable nucleus, the helium ($He^4$) nucleus, can be formed by four-body collisions but the densities at the time of nucleosynthesis make this reaction very improbable. So helium nuclei must form by two-body collisions, the first step of which is formation of deuterium ($D^2$). Deuterium formation is inhibited by a large photon-to-baryon number ratio. So temperature has to be lower before a substantial amount of deuterium is formed without being blasted apart by more energetic photons. But by this time the rate of reaction that converts $D^2$ to $He^4$ lowers. This condition is known as the deuterium bottleneck. The amount of $D^2$ that remains unconverted at the end of nucleosynthesis depends crucially on baryon density at that time as can be seen from the reaction $D^2+p^1 \rightarrow He^3+\gamma$.
If baryon density is low, reaction is slow and $D^2$ is more abundant. Thus, the amount of primordial deuterium places strict bounds on the baryon density of the universe. It is estimated that ordinary baryons form $5 \%$ of the critical density of the universe today. Estimates of total matter density gives $\Omega_m=0.3$ approximately.\cite{weincosmo}
\item A high photon-to-baryon number ratio ensures that at temperatures comparable to nuclear binding energies, a nucleus is destroyed as soon as it is formed by a highly energetic photon. This inhibits production of nuclei till temperature falls well below nuclear binding energy.
There is no stable nucleus with five to eight nucleons. This prevents formation of heavier nuclei.
\item The proton-to-neutron number ratio does not remain a constant of time as the rate of reaction $p^1+e^- \rightarrow n^1+ \nu_e$ falls below that of the decay of neutron when the temperature of universe drops. This explains the greater number of protons in the universe today than neutrons. The photon-to-baryon number ratio, however, remains a constant of time even after the process of nucleosynthesis is completed as processes if photons are created and destroyed in equal proportion during the expansion of the universe. An estimate of the photon number density can be made from the blackbody spectrum, $\int_0^{\infty}{\frac{8\pi\nu^2\,d\nu}{e^{h\nu/kT}-1}}$. At $2.7$ K, this number is around $410$ per cc.
\item High photon-to-baryon number ratio also accounts for why recombination occurs only at $0.25$ eV.
\item With the drop in temperature, reaction rates become smaller and smaller, till finally they dip below the expansion rate of the universe. When this happens, the reacting species fall out of equilibrium and their abundances are frozen to be estimated aeons later by us!

\end{itemize}

\section{A flat universe}
Since nucleosynthesis makes accurate predictions about present-day densities of various elements, its assumptions about conditions in the universe at a redshift of around 1100 can be considered valid. The estimation of Jeans length can then be based on these. From the anisotropies in the CMBR, we know that the largest wavelength acoustic oscillation at the time of last scattering subtends an angle of about $1^{\,\circ}$ in the sky today. If one defines the angular diameter distance $d$ as $d = x/\theta$, then for a flat universe $d = a_0\,\chi = \int_0^z{\frac{dz'}{H(z')}}$. The best fit to the data provided by anisotropies seem to point towards a flat universe.\cite{flat, edjcope} Observed estimates of matter density suggest that $\Omega_m=0.3$ only. This means that for the universe to be flat as implied by anisotropies in the CMBR, there must exist a huge storehouse of energy as yet unaccounted for. This is yet another indication that the universe is now dark energy dominated.

\chapter{AGE OF THE UNIVERSE}
\label{chap6}
\section{The problem}
It is reasonable to expect that a correct theory of the universe gives it an age at least equal to the age of its oldest constituent. The age of the universe depends on which theory of the universe one believes in. For many years people used theology to address this issue. The very earliest attempts at arriving at a date for the creation of universe were calculations that used biblical texts as record of history. This made the universe a few thousand years old. With the advance of science, people from different branches of science started estimating the ages of the sun and earth. The first falsifiable proof of age of the earth came from geologists who used radioactivity to date rocks. In the $19^{th}$ century, Lord Kelvin determined the age of the sun by assuming that the energy radiated by the sun was in fact obtained by gravitational collapse that is, from the potential energy of matter in the star. The more it compressed, the more the heat that was radiated. 
The total potential energy, $P.E.$, of a spherical configuration of matter of uniform density, $\rho$, and radius, $R$, can be calculated as 
\begin{equation}
P.E. = \int_0^R{\frac{G \rho \frac{4 \pi r^3}{3} \rho 4 \pi r^2 dr}{r}} = \frac{3}{5}\frac{GM^2}{R}
\end{equation}
where $M=\frac{4}{3} \pi R^3 \rho$ is the total mass in the system. If the luminosity of sun, $L$, is $3.84*10^{26}$ watts, the radius of the sun, $R$, is $6.95*10^8$ m and the mass of the sun, $M$, is $2*10^{30}$ kg, one obtains that the sun can not be more than $P.E._{sun}/L$ years old. Substituting the said values, Lord Kelvin calculated that the age of the sun is not greater than 25 to 30 million years old. This in turn, meant that the earth was not older than 30 million years old. His estimate, however, was not consistent with those of the geologists who had found much older rocks on earth. He was also at odds with evolutionary biologists who argued that life had existed on earth for at least a few hundred million years before evolving to such diversity as exists today. They studied fossils and rates of sediment deposition to come to their conclusion. Lord Kelvin's argument was, of course, abandoned when it was discovered that the source of sun's energy was nuclear fusion and stellar life cycle was better understood.

A similar situation exists today. The age of the oldest stars in the universe is estimated to be not less than 12 billion years.\cite{sami} But if the universe were thought to contain only dust like matter and radiation, its calculated age is not more than 8-10 billion years. Just as in the case of Lord Kelvin's argument, an independent discovery such as that of the solar nuclear fusion process, is required to remove the current inconsistency. A non-zero value for the cosmological constant could resolve the issue.

\section{Star cluster ages}
Oldest stellar age is calculated by observing globular clusters. A cluster is a group of stars that are believed to have been formed from the same cloud of gas at about the same time. Globular clusters are dense groups of many hundreds of thousands of stars that are gravitationally bound to each other. And as the name suggests, they are spherical clusters of stars and are usually found in the halos of galaxies. They are thought to be some of the oldest objects in the universe for the following reasons. They are generally free of interstellar dust, which is thought to have been accreted to form stars already. From spectroscopic studies, their stars are found to have very low metallicity, which is an indication that their stars formed before heavier elements were synthesised in the universe through stellar processes. 

The age of a star cluster is usually determined in the following manner.
According to stellar evolution theories, the mass of a star determines its absolute luminosity and the time it will take to burn all the hydrogen fuel in its core as well as its ultimate fate. This nicely explains the fact that, in a graph of absolute luminosity of star versus temperature of star, only specific regions are populated by stars. This plot is called the Hertzsprung-Russell diagram (henceforth called H-R diagram) named after the first people to plot such a graph.

To extract information from the H-R diagram of a cluster, one assumes that all stars in a cluster were formed at approximately the same time, and with the same constituents but with varying masses. Then, by determining from the H-R diagram of the cluster the heaviest star still burning hydrogen in its core, or to throw some jargon in, the turn-off point of the main sequence, one may determine the age of the cluster.

The assumption that all stars in a cluster are approximately the same age is validated by the observation that usually, for a cluster, one is able to fit a theoretically predicted isochronic curve to the data in the HR diagram. An isochronic curve plots the luminosity of stars of varying masses against their temperature at a given time after their birth. If stars varied widely in their ages, no single isochronic curve could be fitted.

The advantage in a cluster is that the distances to the stars in it are approximately the same. Their observed radial velocities seem to support this. This means that the apparent magnitudes of various stars in the cluster differ from their absolute magnitudes by the same amount. And thus, one may obtain the H-R diagram  of the cluster simply by plotting the apparent luminosity of the stars against some measure of their temperature. In fact, one may also calculate distance to the cluster by comparing the magnitudes of the main sequence stars in the cluster to those of a nearby cluster whose distance from earth is known, taking into account variations, if any, due to stellar metallicities.

The ages of the oldest star clusters thus dated are around 12 billion years.\cite{ageestimates}

\section{Age of the universe in cosmological model}
To calculate the age of the universe in a theoretical model, one goes back to the relation 
\begin{equation}
dt=-\frac{dz}{H(z)(1+z)}
\end{equation}
Then the age of the universe, $t_0$ is $\int_0^{t_0}{dt}=\int_0^{\infty}{\frac{dz}{H(z)(1+z)}}$. To make things simpler, we again take the case of a flat universe where $\Omega_K = 0$. For a flat matter-filled universe, the age comes out to be $\frac{2}{3 H_0}$, which for $H_0=70$km\,s$^{-1}$/Mpc is about 9 billion years.
\footnote{However, according to Hubble's first estimate, his constant was about $500$km\,s$^{-1}$/Mpc. The reciprocal gives a Hubble time of about 2 billion years. Both 2 and 9 billion years are smaller than the age of the sun, which, according to estimates from stellar lifecycle, is around 10 billion years.}
However, the presence of the cosmic microwave background indicates that our universe also has some of its energy content in the form of radiation. Although its density is now negligibly small in comparison to matter density, there was a time before which radiation was the dominant form of energy. This time corresponds to a redshift of about $z=3000$. To avoid working with complicated integrals involving sums of various powers of $(1+z)$, we calculate separately the time for which universe was radiation dominated and matter dominated. We then add these up to get approximately the age of the universe. The contribution of the radiation-dominated era is $\int_{3000}^{\infty}{\frac{dz}{H_0(1+z)^3}}$ and this is approximately $\frac{10^{-7}}{2 H_0}$ or only a few hundred years. The contribution of the radiation-dominated era is $\int_0^{3000}{\frac{dz}{H_0(1+z)^{5/2}}}$ and this is approximately $\frac{2}{3 H_0}(1-10^{-5})$ or about 9 billion years. Thus one sees that contribution from radiation-dominated era to the age of the universe may be ignored as that era was very short in duration. And we are faced with the problem of the universe being younger than some of its constituents.

\section{A possible solution}
The cosmological constant could be invoked to solve this problem. Using the same argument that was used in the case of the radial coordinate, $\chi$, of a supernova, one sees that a universe with positive $\lambda$ will be older than one that has only matter and radiation content, as at a particular redshift, Hubble's constant is smaller when $\lambda$ is non-zero than when it is.
Another way of increasing the age of the universe in a cosmological model is to allow the universe positive curvature. This also has the effect of lowering Hubble's constant at a particular redshift. But the cosmic background radiation observations have ruled this option out by 
estimating that curvature is very close to zero.

\begin{figure}
\centering
\includegraphics[width=150mm]{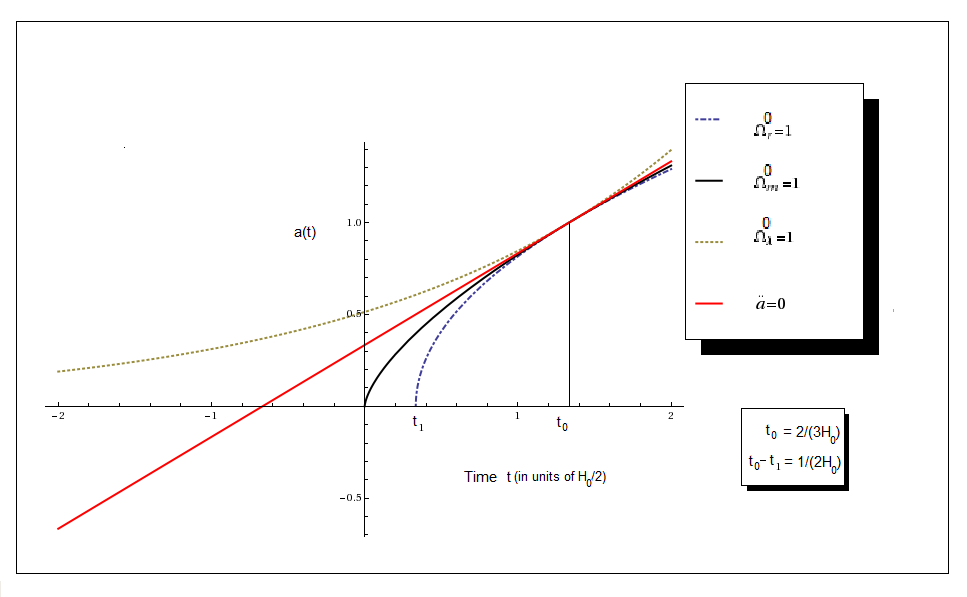}
\caption{Age of the universe in different cosmological models.}
\label{age}
\end{figure}
Figure \ref{age} shows that for a given value of Hubble's constant today, the universe with a greater acceleration is older.

It has now been estimated that for a flat universe containing dark energy and matter, with negligible amount of radiation energy today, cosmological models with $\Omega_{\lambda} \geq 0.5$ make the universe older than 11-12 billion years.\cite{sami} Thus, we have yet another indication that more than half the energy in the universe is still in a form that is a mystery to us.

\chapter{LIGHT ON DARK ENERGY}
\label{chap7}
Thus far in the report we have built up the case for the existence of a non-zero value for the cosmological constant. We have, however, only treated it as a constant in Einstein's equations and not given it any physical significance. This question was raised way back in the 1960's, long before the observation of accelerated expansion of the universe. At that time, the interest in the cosmological constant had been stirred by the need to explain a sudden enhanced deceleration of the expansion of the universe inferred from the observed redshift-luminosity distance relation of quasars. The observations showed a large number of quasars in different stages of evolution clustered around a particular value of redshift, $z$. The cosmological constant was invoked to explain these apparent anomalies.\cite{burbs} In 1968, Zel'dovich \cite{zeldo} interpreted the cosmological constant as vacuum energy. Let us see analyse this interpretation. Although, at first, it is counter-intuitive to think of an energy associated with vacuum, we will assume such energy exists. If so, it is quite sensible to require that all observers measure the same value of vacuum energy density irrespective of their relative motion. If this condition is not imposed, one could distinguish between different states of motion by the value of vacuum energy density measured. Lorentz invariance of energy-momentum tensor of vacuum implies that in a locally inertial frame of reference, it should be proportional to $\eta_{\mu \nu}=Diagonal(1,-1,-1,-1)$, the Minkowski metric. This can be explicitly worked out as follows. If the energy-momentum tensor of vacuum in one reference frame is $T_{\mu \nu}=Diagonal(\rho,p,p,p)$ and is $T'$ in a frame of reference moving at a velocity $v$ in the $x$ direction with respect to the first, the two are related by $T'_{\mu \nu} = \Lambda_{\mu}^{\alpha}\,\Lambda_{\nu}^{\beta}\,T_{\alpha \beta}$ where $\Lambda_{\mu \nu}=\left(
\begin{array}{cccc}
   \gamma & \gamma \beta & 0 & 0\\
   \gamma \beta & \gamma & 0 & 0\\
   0 & 0 & 1 & 0\\
   0 & 0 & 0 & 1
\end{array}
\right)$.

Then,
$T'_{\mu \nu} = \left(
\begin{array}{cccc}
   \gamma^2\,(\rho+\beta^2\,p) & \gamma^2\,\beta(\rho+p)     & 0 & 0\\
   \gamma^2\,\beta(\rho+p)     & \gamma^2\,(\rho+\beta^2\,p) & 0 & 0\\
   0 & 0 & p & 0\\
   0 & 0 & 0 & p
\end{array}
\right)$ 

Equating $T'_{\mu \nu}$ and $T_{\mu \nu}$, one obtains $\gamma^2\,\beta(\rho+p) = 0$ or $\rho = -p$.\\
Thus $T_{\mu \nu}=Diagonal(\rho,-\rho,-\rho,-\rho)=\rho\,\eta_{\mu \nu}$. General covariance then asserts that in general coordinates $T_{\mu \nu} \propto g_{\mu \nu}$.
Thus we see that if vacuum energy were to exist and be included in the total energy content of the universe in Einstein's equations, it would have the form of the cosmological constant.

\section{The origin of vacuum energy}
The idea of vacuum energy finds its origins in quantum mechanics. While traditionally vacuum was understood as empty space devoid of any form of matter or energy, quantum mechanics changed this viewpoint. 

For a particle of mass $m$ in simple harmonic potential $V(x)=m\,\omega^2\,x^2/2$, quantum mechanics says that the lowest energy attainable by the oscillator is equal to $\frac{1}{2} \hbar \omega$ and not zero. This is different from its rest mass energy and the minimum or zero-point energy is a purely quantum mechanical effect. (A classical simple harmonic oscillator at rest in the position of stable equilibrium can be said to have zero kinetic energy.) Similarly, in quantum field theories, the ground state of fields is not that of zero energy. We can thus calculate the energy density of the vacuum by adding up the zero-point energies of all the vibrational modes of the quantum fields we are considering. Vibrational modes with shorter wavelengths have higher frequencies and hence, higher energies. If we assume spacetime is a continuum, the modes are allowed to have arbitrarily short wavelengths. This leads to the vacuum energy density becoming infinitely large. If, however, we impose that spacetime is discrete at a certain finite length scale, then we need consider only those modes of vibrations of fields with wavelengths greater than this length. This will give a finite value for the energy density of vacuum. The cutoff length mentioned is one formed by fundamental constants from quantum mechanics and the special and general theories of relativity and is called the Planck length. It is equal to $\sqrt{\hbar\,G/c^3}$ and has the numerical value of about $1.6 \times 10^{-35}$ metres. Similarly one can define the Planck mass to be $\sqrt{\hbar\,c/G}$ which is numerically equal to $2.2 \times 10^{-8}$ kg. In units of Planck scale, the value of the current estimated value of the energy density associated with the cosmological constant, approximately $10^{-29}$g/cc, is of the order of $10^{-123}$ while the predicted value of vacuum energy density in the same units is of the order of 1. The question of why $\lambda$ is so small and yet not zero, and vastly different from the predicted value of vacuum energy is sometimes called the cosmological constant problem.

\section{Casimir effect}
We can raise the question of whether or not vacuum energy can be detected in any way other than by its gravitational effect. One would imagine that because all experiments measured differences in energy between two states, vacuum energy always cancels out and hence there is no way to measure or detect it. But in 1948, the Dutch theoretical physicist Hendrik Casimir showed that the variation in zero-point energy density as the boundaries of a region of vacuum change result in a force that should be detectable.\cite{casimir} For this he considered as boundaries two parallel conducting uncharged plates separated by a distance $a$. The effect can be seen when one compares the energy density of vacuum for various separations $a$. Let us consider the electromagnetic field in the volume between the two uncharged parallel plates. Homogeneous Maxwell's equations have non-trivial solutions. This means that while the mean values of the electric and magnetic fields, $<E>$ and $<B>$ , in vacuum are zero, there are fluctuations about the mean which give non-zero values for $<E^2>$ and $<B^2>$ which in turn contribute to the energy of the electromagnetic field in the vacuum. The requirement that the component of the electric field parallel to the plates vanishes (there are no net currents on the plates) permit only those standing waves for which the separation is an integral multiple of half their wavelengths. $\lambda_n=2\,a/n$. We immediately see that the difference in energy between two successive modes is inversely proportional to the separation $a$. The smaller the separation, the greater is the energy difference and hence, lower is the energy density. Thus, there is a tendency for the two plates to move closer in order to minimise the energy contained in the volume between them. This tendency or force was first predicted and its value calculated by Casimir.\footnote{It has also been argued by E. M. Lifshitz that the Casimir effect can be repulsive and this has recently been experimentally tested by a group at Harvard university.\cite{repcas}} It is noted that for very high frequencies or very short wavelengths, the presence of the plates does not make much difference. The density of allowed modes between the plates is almost the same as in the absence of plates (or equivalently, when the plates are infinitely far away from each other). The finite contribution to the force comes from longer wavelength modes of oscillation. The force exerted per unit area of the plates has been calculated to be $-\pi^2/240\,a^4$.\cite{casimir, milton} The Casimir effect was first qualitatively confirmed by Marcus Spaarnay.\cite{spaarnay} The force has since been measured by Steve Lamoreaux, Anushree Roy and Umar Mohideen \cite{expts} and others. The Casimir effect, since its first prediction, has found applications in many areas of Physics.\cite{soviet} It has been found to play an important role in explaining hadron structure through the bag model that confines quarks, finds application in super-symmetric field theories and could also help physicists test the validity of ideas about the universe having extra dimensions.

The Casimir effect indirectly confirms that vacuum is not devoid of energy but since it measures a force it only measures a finite difference in energy densities of two vacuum configurations. We can say nothing about the total energy density of vacuum from the Casimir effect. One can, of course, ask if the Planck scale is in fact the correct cutoff in the calculation of vacuum energy. If it is, then we need to find a fundamental reason for why that is the case. We would also need to find a mechanism by which to cancel most of this energy out, except for that one part in $10^{120}$ Planck units which has raised quite a furore in the scientific community in the last decade!

\chapter{CONCLUDING REMARKS}
\label{chap8}
The bulk of this report has dealt with understanding the astounding discoveries about our universe of the past decade in terms of the cosmological constant. There are, as mentioned in the introduction, many other points of view in the scientific community about dark energy. We conclude our report with a brief look at some of these ideas and the motivation behind them. This exercise brings into focus the fact that there is no such thing as a correct theory. A theory is only correct as long as it is not proven wrong! The case with the theories of dark energy now is that we need more data before being able to decide which theories can be ruled out.

\section{Cosmological constant problems}
While the cosmological constant has so far been doing a good job of accounting for observations, it raises many questions. It is mostly in an attempt to answer some of these questions that other theories of dark energy emerge. The most important questions about the cosmological constant concern its value. Why is it positive? Why is it so close to zero and yet not zero? Also, as mentioned in the previous chapter, the size of the cosmological constant presents a problem in its being interpreted as the vacuum energy predicted by quantum field theories.
In this context it must be mentioned that the only theoretical prediction that got the magnitude of the cosmological constant right came from the causal set theory.\cite{sorkin} Causal set theory is an attempt to unify quantum mechanics with gravity into a theory of quantum gravity. It postulates that spacetime is only approximately described by a smooth manifold at scales much larger than the Planck scale and that it is, in fact, discrete near the Planck scale. This is analogous to matter appearing continuous at scales larger than atomic scales. This postulate is however, not unique to this theory. Without getting into the details of the theory, we say that it treats the volume of the universe and the cosmological constant as conjugate variables and binds them in an uncertainty principle of the form  $\Delta V \Delta\lambda \approx \hbar$ . This is similar to the relation between the uncertainties in position and momentum of a particle in quantum mechanics. Taking the volume of the universe roughly as $H^{-4}$ where $H$ is the Hubble's constant, and $\Delta V$ as $V^{1/2}$ yields for the fluctuation of the cosmological constant a value of approximately $H^2$. If the mean value of the cosmological constant is taken to be zero, then this theory comes close to predicting the right magnitude of $\lambda$ today, which is $10^{-120}$ in Planck units.

\begin{figure}
\centering
\includegraphics[width=150mm]{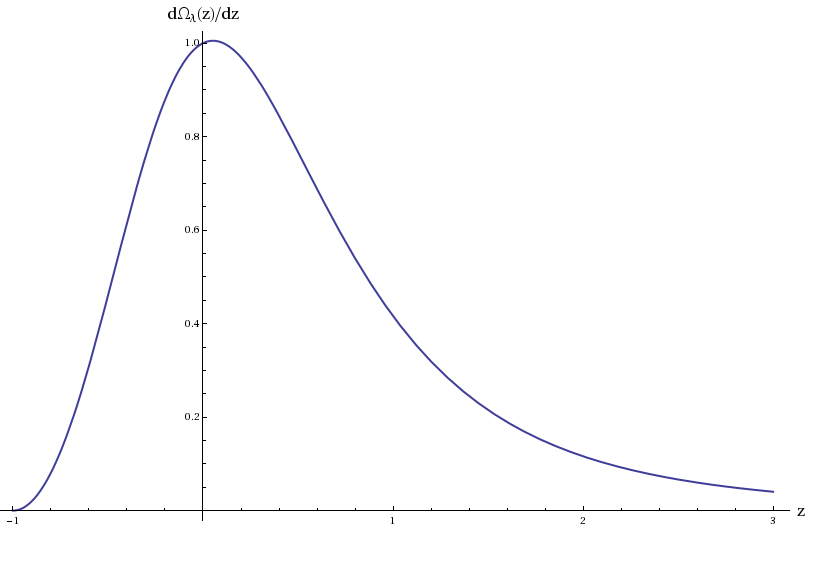}
\caption{Rate of change of fraction of energy density of cosmological constant with redshift versus redshift z.}
\label{domegaz}
\end{figure}

Another group of questions that crops up relates to issues of cosmic coincidence. Accelerated expansion of the universe began at a redshift of around 0.67. Before that time the percentage of dark energy of total energy density was negligible. In the near future dark energy is going to account for almost the entire energy content of the universe. In the currently accepted cosmological model of a spatially flat universe with $\Omega^0_{\lambda}=0.7$ and $\Omega^0_{m}=0.3$ it is seen that $\Omega^{(t)}_{\lambda}$ changes rapidly in a short range of $z$ that is quite close to $z=0$ as seen in figure \ref{domegaz}. In other words, we seem to live in the narrow transitional period between a matter-dominated universe and a dark energy dominated universe. Is this a coincidence? Causal set theory answers this by proposing, as noted above, that the cosmological constant is always of the order of $H^2$.

\begin{figure}
\centering
\includegraphics[width=150mm]{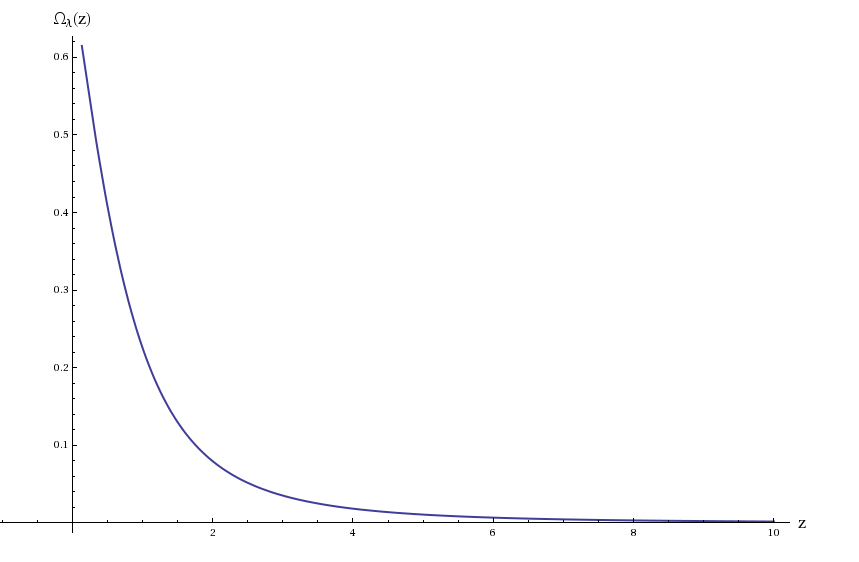}
\caption{Energy density of cosmological constant versus redshift z.}
\label{omegaz} 
\end{figure}

If one tries to avoid the coincidence by assuming that the amount of dark energy in the universe was always close to two-thirds that of critical density, then one could run into trouble with well-established theories of nucleosynthesis and structure formation in the universe apart from possibly not explaining current observations such as CMBR anisotropy and luminosity-redshift relation of distant supernovae. Another set of questions pertaining to coincidences arises from consideration of structure formation. Why do the densities of various energy forms have the values that they do? These values determine when and for how long each of these forms dominates the universe. It is just by chance that dark energy domination began late enough in the lifetime of the universe to allow formation of structures and hence, evolution of life or is there a deeper reason why the universe is the way it is today?

\section{What is dark energy?}
\subsection{A figment of our imagination!}
There is no such thing as dark energy. This point of view arises from considering inhomogeneities in the universe. The accelerated expansion inferred from observations of supernova luminosities and redshifts is not considered to be applicable to the universe as a whole but only to that part of the universe that we observe. The argument is that if we happen to live in an underdense region of an inhomogeneous universe (a void), our region would have a greater local Hubble's constant than in neighbouring overdense regions because greater energy in those regions has caused greater deceleration in the scale factor of that region. Thus, when we look at those regions Hubble's constant is smaller and it appears as though the universe is accelerating. For this explanation to be correct we need to be residing in a very large void. Observations at large scales so far suggest that the universe is homogeneous. This belief is further strengthened by the almost isotropic CMBR. Also, the strong energy condition, which all forms of energy except dark energy obey, prohibits the void from expanding with an acceleration.

\subsection{Quintessence} 
Even when there is consensus among scientists about the existence of dark energy, there is a lot of variety to be found in the opinions held by people about what exactly dark energy is. 
One such idea is that of quintessence which was proposed in 1998 by R. R. Caldwell, R. Dave, and P. J. Steinhardt.\cite{quint} The motivation for its introduction was derived from some of the problems with the cosmological constant mentioned above. It is also natural to move from the cosmological constant that is characterised by a single number, its magnitude, to other theories which allow the energy density and equation of state of dark energy to vary over time and space. The condition on the value of $w$, which appears in the equation of state, obtained in Chapter 3, where we considered the case of a universe containing only dark energy, for positive acceleration was that $w \leq -1/3$. A tighter bound on the equation of state of dark energy can be obtained as follows. From independent sources we know that the universe is spatially flat (CMBR) and has $\Omega_m=0.3$ (estimates of baryon density and dark matter density from galaxies and interstellar matter). \footnote{The observational estimation of matter density assumes that galaxy clusters, the largest structures in the universe, are representative of the relative proportions of dark matter and baryons in the universe.} We also know that the energy density of radiation is negligible. This implies that around $70 \%$ of energy density is contributed by dark energy. Plugging these values into Einstein's equation for the acceleration of the scale factor and allowing $w$ of dark energy to vary, one obtains for positive acceleration the condition that $w \leq -1/2$.

Quintessence is a dynamic dark energy in the sense that it generally has a density and equation of state that varies through time and space. The quantity $w$ is allowed to vary anywhere between $-1/3$ and $-1$. Quintessence is a scalar field that can be visualised as a set of springs spread over all space, each stretched differently. In this picture Einstein's cosmological constant would correspond to springs of uniform spring constant stretched to the same length and being motionless indicative of constancy through both space and time.
In the simplest picture, the scalar field $\phi$ is characterised by energy density $\rho_{\phi} = \dot{\phi}^2/2 + V_{\phi}$ and pressure $p_{\phi} = \dot{\phi}^2/2 - V_{\phi}$ where V is the potential associated with the field. Then, from ($\ref{rhodot}$), one obtains for the conservation of energy the equation $\ddot{\phi} + 3\,H\,\dot{\phi} = -dV/d\phi$. Work is being done to distinguish between different models of quintessence that are characterised by different values of $\dot{w}$, that is, by how the equation of state of the scalar field is varying with time.\cite{caldwell} Work is also being done to segregate quintessential effects from that of a cosmological constant.
It is possible to distinguish between a time-varying and constant dark energy by measuring the acceleration of the universe at different times. There are missions such as NASA's Joint Dark Energy Mission being planned for the same. 

A dynamic dark energy is also expected to have different implications for fundamental physics from those of a cosmological constant.\cite{stein} The value of the cosmological constant remains unchanged throughout the history of the universe. As it was set in the very early universe, it bears the imprints of physics near the Planck scale. But quintessence, because of its dynamic nature, could also contain vital information about low energy physics.

\subsection{$k$-essence}
$k$-essence is also a scalar field model of dark energy. It seeks to explain the coincidences in a slightly better way than by saying that they are mere outcomes of chance.\cite{kessence} It also does not resort to anthropic reasoning. This theory says that $k$-essence begins to have negative pressure only after the era of matter-radiation equality. This means that it can overtake the energy density of matter only after matter energy has dominated the universe for a while. This implies that cosmic acceleration begins only after enough time has been allowed in the universe for structure formation.

\subsection{Phantom energy}
Phantom energy is a type of energy that has $w < -1$. It violates not only the strong energy condition, but also the weak and dominant energy conditions. We recall that the essence of the dominant energy condition is that nothing can travel faster than the speed of light. While this makes phantom energy seem unphysical, models of phantom energy are also being proposed that allow for subluminal speeds for sound waves through the medium.\cite{nevinwein} According to Einstein's equations, phantom energy density actually increases in an expanding universe. This leads to increasing acceleration eventually leading to what is called the ``big rip'', a scenario in which in the distant future all structures and eventually even subatomic particles are torn apart due to the repulsion caused by increasing negative pressure.

\section{The road ahead}
Since 1998, there has been a large amount of evidence piling up in support of the existence of a mysterious form of energy found to fuel cosmic acceleration. But the jury is still out on the question of how to explain the observations best. It is evident that the answer holds the key to understanding some deep and fundamental truths about our universe. At present there is a lot of room to explore many different models of dark energy. We need more data so as to place stricter bounds on the values of observable quantities and narrow down the list of candidates for dark energy. Technological advances and sophisticated instrumentation promise to help the cause. 
There sure is a bright future for studies in dark energy!

\appendix
\addcontentsline{toc}{chapter}{APPENDICES}
\chapter{Olbers' Paradox}
\label{appa}
Olbers' paradox asks why the night sky is not much brighter than it appears now if certain assumptions were made about the universe. 
The question has been contemplated over the years by Heinrich Olbers, after whom it is named, Johannes Kepler of the Kepler's laws of planetary motion fame,  Halley of the comet fame and Cheseaux, a Swiss astronomer, and more recently Hermann Bondi among many others.
The assumptions are that the universe is homogeneous, isotropic, static, eternal, infinite and Euclidean, transparent and that the number of stars per unit volume of the universe is finite and fixed, say $n$. Of course, when one says the universe is homogeneous and isotropic one means that it is so to a good approximation at large enough distance scales. This means that at any given time the stars are all uniformly distributed through space, and the luminosity, $L_0$, of a star is constant through space and time. One could then ask how much light the earth receives per unit area per unit time from stars in the universe. The flux, $F$, of radiation energy received at earth from stars in a shell of radius $r$ and thickness $dr$ is proportional directly to the volume of the shell and inversely to the area of the sphere formed with the source as centre and distance to earth as radius. Thus as distance to a shell increases, fall in radiation intensity with square of distance is compensated for by the increase in the number of stars owing to the increase in the volume of the shell. Hence the total flux of radiation energy that the earth receives is simply proportional to $\int_0^{\infty}{dr}$.
\begin{equation}
F = L_0 \int_0^{\infty}{\frac{n\, 4 \pi\, r^2\, dr}{4 \pi\, r^2}} = L_0 n \int_0^{\infty}{dr}
\label{flux}
\end{equation}
This means that sky ought to be infinitely bright! But wait, in which direction and when? In the kind of universe imagined above, an observer cannot look in a direction without seeing a star. Uniform random distribution of stars through the universe assures every direction in space has an equal chance of containing a star. And the infinite age of the universe allows one to look at light from any point however far it be. Thus if one looks far enough one will eventually sight a star. So sky ought to seem infinitely bright in every direction. And since the universe is also assumed to be static and the number of stars per unit volume unchanging with time, this has to be true always. That is one long day for the observer! But luckily, an observer in the real world does not have to suffer such a long, hot day. However, he/she now has the task of explaining such luck.

Over the years, people have tried to resolve this paradox in many ways.
One proposed solution was that the universe was not transparent to radiation and that interstellar matter blocked stellar light from reaching the observer. But the laws of thermodynamics show this did not resolve the issue. If the interstellar material is colder than the star, radiation will be absorbed. And because the universe is static and eternal, the material has had time enough to absorb radiation, reach a state of equilibrium with the radiation and to begin radiating at the temperature of the star. Hence, to the observer it does not matter whether it is the star that radiates or the interstellar matter, and the paradox persists. The same is the case if one accounts for the fact that starlight from stars could be blocked by other stars in the same line of sight. Since the universe has had an eternity to equilibrate, one would expect the sky to glow uniformly at the temperature of a star.

Another set of people has claimed that the paradox holds only if each star is individually said to have existed forever and that the finiteness of the age of stars resolves the paradox. But this is not, in fact, the case. For the finiteness of the age of a star does not change the fact that every direction the observer looks in contains a star. Every direction still has an equal probability of containing a star as the distribution of stars is assumed to have been uniform in the past too. One also notes that the integral in (\ref{flux}) remains divergent because of the assumption that the number of stars per unit volume remains the same through time although the locations of individual stars vary with their birth and death.

Yet another explanation of why a large part of the night sky appears dark could be that stars are not distributed uniformly through space. They could be distributed such that they are clumped up in some parts of the universe, leaving large empty spaces elsewhere. A fractal distribution of stars has been proposed by Mandelbrot and others. However, observations suggest that at very large scales, our universe is uniform.

One try at resolving the paradox is by postulating that the universe has only a finite number of stars in it. While this definitely means that the flux reaching an observer on earth is finite, it still does not explain why the night sky is as dark as it is and as cool, because the number of stars in the universe now is believed to be of the order of $10^{21}$.

One could try resolving the paradox by altering the geometry or topology or both of the universe, retaining the other assumptions. Let us consider an altered topology first. The universe can be imagined to be bounded and finite, like a cube. However, because there is nothing `outside' of the universe, one must impose the condition that when one tries to look beyond any one face of the cube, one ends up looking through the face opposite to it. When the opposite faces of the cube are identified thus, universe becomes equivalent to an infinite lattice with the cube as the unit cell. This brings one back to the paradox. The problem mentioned with assuming a cube universe, or in general, a box universe, is better understood if one relates it to the experience of living on the surface of the earth. If one were to start moving straight ahead from a point on the surface of the earth in some general direction, he/she would eventually come back to the starting point. Of course, the motion is restricted to only the surface of the sphere. A non-Euclidean universe, with negative or positive curvature, also does not resolve the paradox as it either extends to infinity or curves upon itself, both of which scenarios have been considered above.

\begin{figure}
\centering
\includegraphics[width=85mm]{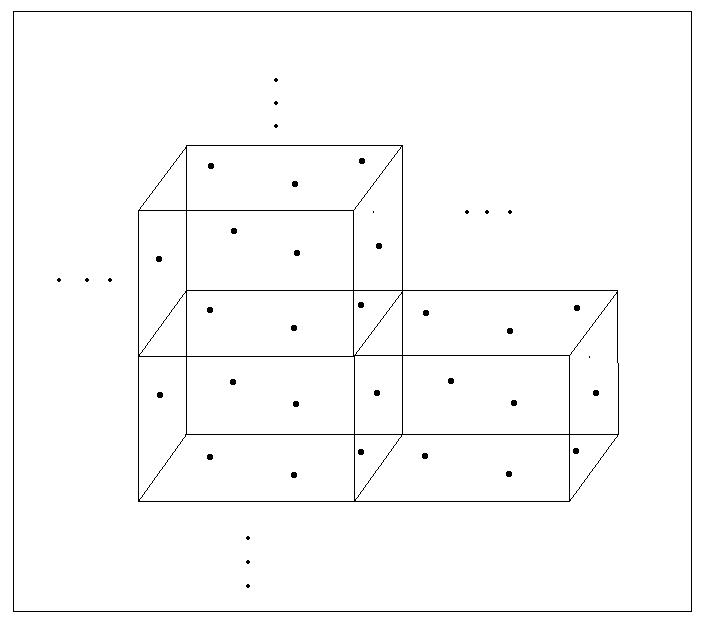}
\caption{\label{fig2}Identifying opposite faces of a box universe makes it equivalent to an infinite universe.}
\end{figure}

To resolve the paradox, some people were prompted to think that maybe the light reaching the observer came from a sphere of only a finite radius. Taking to be true the finiteness of the speed of light, this meant one of two things. Either the universe came into existence at a finite time in the past, not allowing light from great distances time enough to reach the observer or all the stars in the universe `turned on' at a finite time in the past. This immediately makes the integral in (\ref{flux}) finite. Both these suggest that the universe is not eternal.

A plausible solution presented itself when Hubble's observations seemed to indicate that the universe was not static, but in fact, expanding. In such a universe, $L_0$ no longer remains constant, but changes due to redshift effects, implying that stars further away from the observer contribute less to the flux he/she receives. 

In the Big Bang cosmology, both the finite age of the universe and its expansion aid in avoiding the paradox. By some simple calculations one can see which effect contributes more towards resolving the paradox. If one assumes a static universe of a finite age, say $T$, (\ref{flux}) becomes
\begin{equation}
F = L_0 \int_0^{cT}{dr} = L_0\,n\,c\,T
\end{equation}
If the effects of expansion are also included, and universe is assumed to be matter-dominated, flux is given by
\begin{eqnarray*}
F &=& L_0\, n \int_0^{\infty}{\frac{dz}{H(z)(1+z)^2}}  \\
  &=& \frac{2}{5} \frac{L_0\, n\, c }{H_0} \\
  &=& \frac{3}{5} L_0\, n\, c\, T
\end{eqnarray*}

Here the age of the matter-dominated universe is $T=\frac{2}{3H_0}$.
Thus, one sees that the added effect of expansion on a finitely old universe is to reduce flux reaching the observer by a factor of about two.
This shows that the finite age of the universe has the more prominent effect in resolving Olbers' paradox than redshifting effects in the Big Bang cosmology.
It must be noted that redshift effects alone can also make the flux reaching an observer finite, even if the universe is assumed to be infinitely old. In an eternal universe like the deSitter universe, one obtains $F = \frac{L_0\, n\, c }{H_0}$. In steady state cosmology, where the Big Bang is absent, and the universe is considered to be eternal, it is this argument that finds a way out of an infinite flux reaching the observer. The steady state theory has lost its popularity among cosmologists owing to the growth in evidence for a Big Bang universe. \footnote{Reference: Ray D'Inverno, Introducing Einstein's Relativity, Oxford, 1992.}

\chapter{Symmetric spaces and Killing vectors}
\label{appb}
\section{Symmetric spaces}
The question is, given the symmetries of a metric space, can one construct its metric? In cosmology one is usually interested in homogeneous and isotropic spaces. Before tackling the problem in a mathematically systematic way, we try to get a visual and intuitive picture of the same. By homogeneity, we mean the idea that space is everywhere the same and by isotropy we mean that space looks the same in every direction from a point. We start by imagining a homogeneous and isotropic space in one dimension. An example of this is an infinitely extending straight line. A line-segment will not do as it distinguishes its end-points from the other points on the line. Another example is the set of points on any arbitrary closed or open curve. A creature confined to the curve cannot distinguish one point from the other although someone who sees the curve drawn on a sheet of paper or wiggled in space can immediately see that the curve `curves' differently at different points. Coming to two dimensions, an infinite plane is an easy and correct guess at a homogeneous and isotropic space. Another example is the surface of a sphere. In this case, any arbitrary closed surface will not do because a being confined to the surface can distinguish one patch from another by drawing a triangle on the surface and checking what the three angles add up to. Formally it is said that one-dimensional curves have no intrinsic curvature but higher-dimensional objects have intrinsic curvature. An intrinsic property of a space is one that a being on the space can figure out without having to observe his abode from a higher-dimensional space. Coming back to adding up sums of angles of a triangle on a surface, it is noticed that the sum is greater than $180^{\,\circ}$ on a sphere. It is also possible that the sum is less than $180^{\circ}$ for a surface. Such a surface is said to be negatively curved while surfaces on which the sum is greater than $180^{\circ}$ are said to be positively curved. Surfaces with uniform negative curvature are called hyperbolic surfaces. There is, of course, also the possibility that a surface has different curvatures at different points, but we are not interested in such surfaces right now.
Moving on to three dimensions and taking the cue from the examples in two dimensions we say that there are three different types of isotropic and homogeneous three-dimensional spaces corresponding to constant positive, negative and zero curvature. The last of these is simply the Euclidean space that we are most familiar with and can most easily visualise. The other two can be regarded as higher-dimensional extensions of spherical and hyperbolic surfaces. Having listed down these spaces, our task now is to write down a metric each for these spaces. Once again we start from lower dimensions, and consider spaces other than flat spaces whose line element is given by $ds^2 = \Sigma_i dx_i^2$ where the space is $i$-dimensional. We look at positively curved spaces that are embedded in higher-dimensional Euclidean spaces.

Let us take the circle in two dimensions. It has the equation $x^2 + y^2 = a^2$. This implies that $xdx = \mp ydy$ and the metric $ds^2 = dx^2 + dy^2$ can be rewritten as 
\begin{equation}
ds^2 = dx^2 + \frac{x^2 dx^2}{a^2 - x^2}
\end{equation}
When the coordinates are transformed as $x'=\frac{x}{a}$ and $y'=\frac{y}{a}$, the metric takes the form $ds^2 = a^2[dx'^2 + \frac{x'^2 dx'^2}{1 - x'^2}]$.
Similarly the metric of an $n$-dimensional spherical surface embedded in an $(n+1)$-dimensional Euclidean space can be written as 
\begin{equation}
ds^2 = a^2[dx^2 + \frac{(x.dx)^2}{1 - x^2}]
\label{sph}
\end{equation}
where $x$ is an $n$-dimensional vector.

This is as far as our imagination will take us. We now make the promised systematic mathematical study of symmetric spaces.

\section{Killing vectors}

Choice of coordinate system can obscure symmetry. Hence the need for a coordinate independent description of symmetries of spaces. Consider the metric on the surface of a sphere. The line element, in spherical polar coordinates reads
$d\tau^2 = r^2[d\theta^2 + sin^2\theta d\phi^2]$
It looks as though the metric is dependent on the $\theta$ coordinate, while in fact, one knows that the metric can be made coordinate-independent with the transformation

\begin{eqnarray*}
x &=& r \, sin\theta \, cos\phi \\
y &=& r \, sin\theta \, sin\phi \\
z &=& r \, cos\theta
\end{eqnarray*}

and the line element looks like $d\tau^2 = dx^2 + dy^2 + dz^2$. Of course, in the second set of coordinates, we see the spherical surface as embedded in a 3-dimensional Euclidean space, and hence use three coordinates to represent a point on the sphere where the sphere is the surface $x^2 + y^2 + z^2 = r^2$.
As an example from cosmology, consider the metric of the deSitter universe where $H$ is a constant. 
\begin{equation}
d\tau^2 = c^2 dt^2 - e^{2Ht}[dr^2 + r^2(d\theta^2 + sin^2\theta \, d\phi^2)]
\label{desitter}
\end{equation}

Written in this form, the metric seems eternal. But with the following coordinate transformation, it, surprisingly, looks static.
\begin{eqnarray*}
R &=& e^{Ht}\,r \\
T &=& t - \frac{1}{2H} \log{1-\frac{r^2\, H^2\, e^{2Ht}}{c^2}}
\label{desitterctrans}
\end{eqnarray*}
The metric now reads
\begin{equation}
d\tau^2 = c^2 (1-\frac{H^2\,R^2}{c^2}) dT^2 - \frac{dR^2}{(1-\frac{H^2\,R^2}{c^2})} - R^2[d\theta^2 + sin^2\theta \, d\phi^2]
\end{equation}

We now precisely define the symmetries homogeneity and isotropy.

{\it A metric space is homogeneous if there exist infinitesimal isometries that carry any one point to any other point in its immediate neighbourhood.} 

{\it A metric space is said to be isotropic about a point X if there exist infinitesimal isometries that leave X fixed and carry any direction to any other direction.}

But, to make complete sense of the above definitions and to check if they are consistent with the meaning we attributed to these terms at the start of the chapter, we need to know what an isometry is. An isometry is a coordinate transformation $x \rightarrow x'$ that leaves the functional form of the metric unchanged, that is, if $g(x)=x^2 + 2x$ is the metric in one coordinate system, then $g'(x')=x'^2 + 2x'$ is the metric in the transformed coordinate system if the coordinate transformation is an isometry. Or, in short, $g'(x)=g(x)$. One must be careful not to confuse isometry with the definition of a scalar field. For scalars, the value of the function at a point, and not necessarily the functional form of the field, must be the same in all coordinate systems, that is, $s'(x')=s(x)$ where $x \rightarrow x'$.

For infinitesimal transformations of the form $x'^{\mu}=x^{\mu} + \epsilon\,\eta^{\mu}(x)$ where $|\epsilon|<<1$, the condition for isometry is 
\begin{eqnarray*}
g_{\mu\nu}(x) &=& \frac{dx'^{\alpha}}{dx^{\mu}}\!\frac{dx'^{\beta}}{dx^{\nu}}\!g'_{\alpha \beta}(x') \\
              &=& \frac{dx'^{\alpha}}{dx^{\mu}}\!\frac{dx'^{\beta}}{dx^{\nu}}\!g'_{\alpha \beta}(x')
\end{eqnarray*}
When this expression is expanded retaining only terms up to first order in $\epsilon$, one obtains the condition
\begin{equation}
\eta_{\mu;\nu} + \eta_{\nu;\mu} = 0
\label{killing}
\end{equation}
where $\eta_{\mu}=g_{\mu \nu} \eta^{\nu}$.

Any four vector field, $\eta(x)$ that satisfies the above equation is called a Killing vector of the metric. One can thus immediately convert the above definitions of isotropy and homogeneity into conditions on the existence of suitable Killing vectors of the metric.

Homogeneity implies existence of a Killing vector $\eta(x)$ such that at any point $X$, $\eta(X)$ is allowed by the metric to take all possible values. 

Isotropy about a point $X$ requires that there exist a Killing vector $\eta(x)$ such that $\eta(X) = 0$, and for which the first derivatives take all possible values.

We are now convinced that the above definitions capture what we mean by the terms homogeneity and isotropy. Furthermore, we have discovered that symmetries of a metric space can equivalently be talked about in terms of its Killing vectors.

Now we are interested in finding out what is the maximum number of Killing vectors that a metric space can have. 
In cosmology we want to model the universe as a space with no asymmetry. It is this idea that led us to a homogeneous and isotropic universe in the first place. Now we seek to verify if a homogeneous and isotropic space is, in fact, the most symmetric space or not.

It should be noted that any linear combination of Killing vectors is a Killing vector. It can also be shown by differentiating (\ref{killing}) and using properties of the curvature tensor that $\eta_{\mu;\nu;\rho}=-R_{\rho;\nu;\mu}^{\lambda}\,\eta_{\lambda}$.
Thus, a Killing vector at a point, $\eta_{\mu}$, and its covariant derivative at that point, $\eta_{\mu;\nu}$, are sufficient to determine all higher order differentials of $\eta_{\mu}$ at that point. The Killing vector at any arbitrary point can be determined if it is written down as a Taylor series about some point where $\eta_{\mu}$ and $\eta_{\mu;\nu}$ are known.
In essence the $n^{th}$ Killing vector of a metric can be written as 
\begin{equation}
\eta_{\mu}(x) = A_{\mu}^{\nu}(x,X) \, \eta_{\nu}(X) + B_{\mu}^{\rho\,\sigma}(x,X) \, \eta_{\rho;\sigma}(X)
\end{equation}
For an $N$-dimensional space, one must specify $N$ independent values of $\eta_{\nu}(X)$ and $N(N-1)/2$ independent values of $\eta_{\rho;\sigma}(X)$. The latter number is arrived at by recalling that $\eta_{\rho;\sigma}(x)$ is an anti-symmetric tensor as per (\ref{killing}). The values that need to be specified can be considered like coordinates in an $N(N+1)/2$-dimensional space. A set of Killing vectors is said to be independent if they do not satisfy the following equation for constant coefficients $c_n$, $\sum_{n}{c_n \, \eta_{\rho}^n(x)} = 0$. It means that one has at most $N(N+1)/2$ independent Killing vectors for an $N$-dimensional space.

A metric space that allows the existence of the maximum possible number of Killing vectors is called a maximally symmetric space. A homogeneous and isotropic space is maximally symmetric and vice-versa.
A space is homogeneous if it allows $N$ Killing vectors of the form $\eta_{\mu}^{\nu}(x,X)$ with $\eta_{\mu}^{n}(X,X)=\delta_{\mu}^{n}$.
A space is isotropic about point $X$ if it allows Killing vectors at $X$ no freedom to take any value other than zero but allows their covariant derivatives to take all possible values. There can be $N(N-1)/2$ independent Killing vectors. Let us call these $\eta_{\mu}^{lm}$. Then, $\eta_{\mu}^{lm}(x,X)=-\eta_{\mu}^{ml}(x,X)$, $\eta_{\mu}^{lm}(X)=0$ and $\eta_{\mu;\nu}^{lm}(X,X)=\delta_{\mu}^{l}\, \delta_{\nu}^{m} - \delta_{\nu}^{l}\,\delta_{\mu}^{m}$.
Now these $N(N+1)/2$ Killing vectors can be shown to be independent. Thus, a homogeneous and isotropic space is maximally symmetric.

In addition, a metric space that is isotropic about every point must necessarily be homogeneous. This is easy enough to reason out. Any two points in the space can be made to lie on a sphere centred on a suitable point. Isotropy about the centre implies that these two points are equivalent. This scheme can be extended to all the points in the universe to yield that every point is equivalent to every other. Thus a metric space that is isotropic about every point is maximally symmetric.

An obvious question is, how many maximally symmetric spaces can one construct in $N$ dimensions? In other words, how does one distinguish between two different maximally symmetric spaces of the same dimension? It turns out that our hunch in the earlier part of the chapter was right, that a maximally symmetric space is uniquely identified by its curvature, provided the number of positive and negative eigenvalues of the metric are specified. That is, {\it two maximally symmetric Riemannian metrics with the same curvature constant K are locally symmetric.}

We do not look to prove this theorem but just outline where the quantity $K$ comes from. For a maximally symmetric space, it can be shown that $R^{\alpha}_{\alpha}$ is a constant in space. This can be used to define a quantity called the curvature $K$ by $R^{\alpha}_{\alpha} = -N(N-1)K$. And now we take that the theorem is correct.

The importance of the theorem is in the fact that it tells us there is one quantity that uniquely identifies a maximally symmetric metric space and when scaled, there are only three possibilities for this quantity depending on whether it is zero, positive or negative. Thus we can construct maximally symmetric spaces by any method in any coordinate system we like only ensuring that we have accounted for three values of $K$.

To conclude our study we construct maximally symmetric metrics in $N$ dimensions in the same intuitive way we had done earlier, but this time accounting for spaces with negative curvature as well.
In an $(N+1)$-dimensional space with the following metric,
\begin{equation}
d\tau^2 = K\,A_{\mu \nu} dx^{\mu} dx^{\nu} + dz^2
\end{equation}
we embed an $N$-dimensional pseudosphere whose equation is $A_{\mu \nu} x^{\mu} x^{\nu} + K^{-1}\,z^2 = 1$.
The metric on the surface is then given by
\begin{equation}
d\tau^2 = K\,A_{\mu \nu} dx^{\mu} dx^{\nu} + K\,\frac{[A_{\mu \nu} x^{\mu} dx^{\nu}]^2}{1-A_{\alpha \beta} x^{\alpha} x^{\beta}}
\label{genmetric}
\end{equation}

A two-dimensional negatively curved surface can be constructed using the above method.
Consider the surface $x^2+y^2-z^2=1$ embedded in a three-dimensional non-Euclidean space with line-element $ds^2=dz^2-dx^2-dy^2$.
Substituting for $dz$ using $z\,dz=x\,dx+y\,dy$ and the equation of the surface, we obtain the line-element on the surface to be
\begin{equation}
ds^2=-dx^2-dy^2-\frac{(x\,dx+y\,dy)^2}{1-x^2-y^2}
\end{equation}
The above equation is in the form given in (\ref{genmetric}) with $K=-1$. A three-dimensional negatively curved surface can be constructed by considering the surface $x^2+y^2+u^2-z^2=1$ embedded in a four-dimensional non-Euclidean space with line-element $ds^2=dz^2-dx^2-dy^2-du^2$. 
Once a homogeneous and isotropic three-dimensional surface has been constructed, it can be used to construct $(3+1)$-dimensional spacetime with Lorentzian signature, the metric of which will be given by
\begin{equation}
d\tau^2 = dt^2 - a^2(t)\,ds^2
\end{equation}
where $ds^2$ is the line-element of the said surface.

A general method for constructing the metric of a surface embedded in a higher dimensional space is the following. Consider a surface $g_{ab}\,x^{a}\,x^{b}=1$ embedded in a space with metric $g_{ab}$. The unit normals to this surface are given by $n_c=\nabla_c(g_{ab}\,x^{a}\,x^{b})/2 = g_{ac}\,x^{a}$. The metric on the surface will then be $h_{ab}=g_{ab}-n_a\,n_b$. That is, components that stick out normal to the surface are eliminated.

Let us try constructing Killing vector fields for the simplest surface, a flat two-dimensional Euclidean space. From our result above we know that this surface must have three independent Killing vector fields or in other words three symmetries. We also know that these correspond to two symmetries of translation, in two mutually perpendicular directions, and one of rotation. The Killing vector fields for translation are $\delta_{n}^{\mu}$ where $n$ stands for the number of the Killing vector field and $\mu$ for the coordinates. Written as two-tuples, these would be (1,0) and (0,1) for translation along each perpendicular direction. For an infinitesimal rotation, the Killing vector field is given by $(x_2,-x_1)$ where the coordinates are $x=(x_1,x_2)$. The three fields can be seen to satisfy our condition for independence. \footnote{Reference: Steven Weinberg, Gravitation and Cosmology, John Wiley \& Sons Inc., New York, 1972.}

\chapter{Cosmic distance ladder}
\label{appc}
Many methods are used to determine astronomical distances. Each of these is effective at different distance scales. The ones that reach out further in space depend on distance measurement schemes at smaller scales. Hence, the analogy to a ladder. 
In the following pages we will have a look at some of the methods of determining distances, some of which truly make one marvel at the ingenuity of their creators. This section is intended only as a brief overview of some of the methods astronomers use to measure distances. The methods have been chosen to give a flavour of the wide variety of approaches taken by the astronomers.

\section{Trigonometric Parallax}
The revolution of earth around the sun causes an apparent shift in position of nearby stars when compared with faraway background stars. If the angular shift in position is $2 \theta$ radians, the mean distance of earth from sun is $d_E$ and the distance from the sun of the star whose shift in position is being observed is $d$, then $d = \frac{d_E}{\theta}$. When $\theta$ is measured in arc-seconds and $d_E$ is taken to be unity, $d$ is obtained in units of parsecs. Thus, for $d_E = 1.5*10^8$ km, 1 pc $\approx 3.2$ light years. Due to limitations on the angular resolution that is possible from earth, distances to stars up to around 30 pc can be measured from earth. However, much better angular resolution is possible from satellites in space which are not clouded by earth's atmosphere. The Hipparcos satellite has catalogued many stars at distances greater than 30 pc, some at 100 pc and above.
It is important to note that in the above calculation we assume that the sun and star are not moving with a transverse velocity with respect to each other.

\begin{figure}
\centering
\includegraphics[width=150mm]{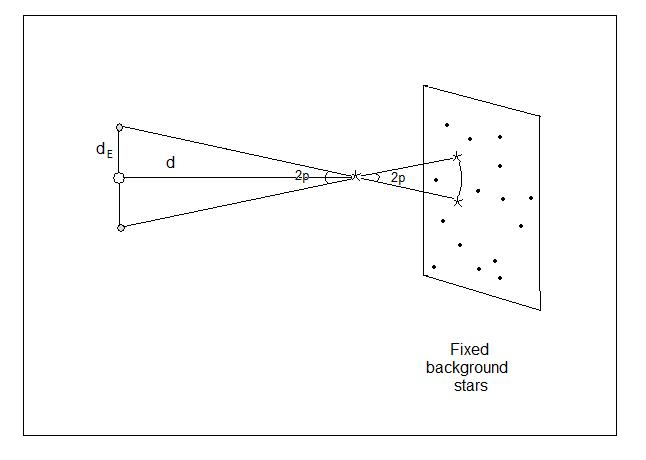}
\caption{\label{trigpar} Trigonometric parallax method of distance estimation.}
\end{figure}

Trigonometric parallax, had it been effective over scales much larger than our galaxy, will help determine curvature of universe when plotted against redshift. But limits on the angular resolution possible with current technology and lack of sufficiently bright luminous objects make this method of distance estimation effective only at smaller than galactic scales.
Currently the Chandra X-ray Observatory in space has an angular resolution of $0.5$ arc-seconds and the Hubble space telescope $0.05$ arc-seconds. Ground based telescopes have poorer angular resolution, the best being around $0.4$ arc-seconds, because of the turbulent effects of atmosphere. They also cannot observe in the X-ray and ultraviolet region of the electromagnetic spectrum owing to absorption by atmosphere.

\section{Moving cluster method}
If stars in a cluster have parallel velocities, they appear to converge at a distant point just as a pair of parallel lines straight ahead of us seem to converge at a distance. This is because while the perpendicular distance between them remains the same, the angle that it subtends at our eye become smaller and smaller until, at last, it is no longer resolvable. The angle formed by the cluster and the convergent point at our eye, say $\theta$, is the same as the angle between the actual velocity of a star in the cluster and its component along our line of sight. This gives the relation between the radial and tangential components of velocity, $v_r$ and $v_t$ respectively, as $v_t = v_r cot\theta$.
\begin{figure}
\centering
\includegraphics[width=150mm]{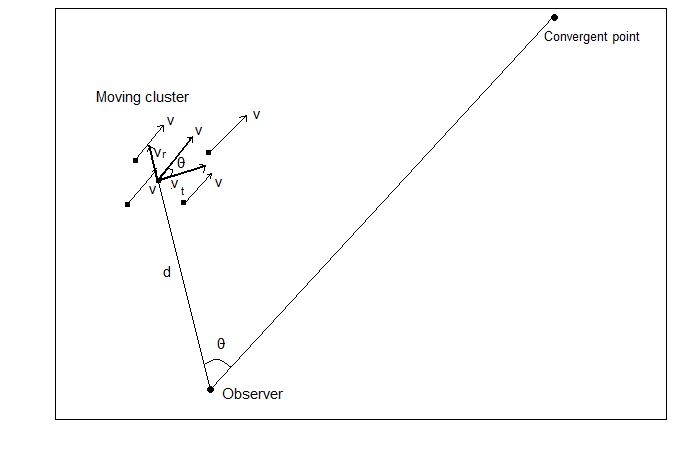}
\caption{\label{movclus} Moving cluster method of distance estimation.}
\end{figure}
$\theta$ is easily obtained by monitoring the motion of stars in a cluster over a period of time. $v_r$ is obtained from the redshift of the spectrum of the star.
For close enough clusters, $v_t = d\,\omega$ where $\omega$ is the angular change in position of star across the observer's line of sight in some unit time and $d$ the distance to the star. $\omega$ when measured in arc seconds per year is called proper motion of the star. Thus, one obtains the distance to the cluster by averaging over distances to many stars in the cluster. Distances to nearby open clusters such as Hyades and Pleiades have been determined in this fashion to be about 41 pc and 135 pc respectively. To make use of this method, stars in a cluster need to be clearly resolved.

\section{Statistical parallax}
Another clever approach to determining distances uses statistics. The idea is that a sample of stars which are believed to be at approximately the same distance have the same distribution of radial velocities as well as transverse velocities. Their radial velocities are 
catalogued by redshift studies and their proper motions are recorded. The above requirement then gives an estimate of the distance to the sample of stars. The risk, of course, is in the assumption made about the nature of the distribution of velocities. This method is also used when the observed stars are at different distances from the observer but the relative distances between each of them is known. Thus, determining one unknown, say the distance to any one star, will determine distances to all the other stars being observed. Hence, one can use stars that are known to have the same intrinsic brightness but are at different distances in this method.

\section{Main sequence photometry}
The Hertzsprung-Russell plot of absolute luminosity of a star versus its temperature is a very useful tool to measure astronomical distances. The important feature of this plot that allows its use as a distance indicator is the fact that depending on its mass, a star, in its lifetime, moves along a specific curve on this diagram. That is to say, the diagram is not randomly populated by stars in all regions. A given group of stars that are all the same age, forms a specific pattern on the diagram, portions of which can be identified with different stages in the life of a star. These patterns do not depend on how far the group of stars is. The distance only determines the apparent luminosities of the stars. Thus, if in the diagram apparent luminosities are plotted instead of absolute luminosities, the relative distances between different groups can be estimated by matching the patterns, in particular, the branch of the pattern called the main sequence. This branch refers to the hydrogen-fusing stage in a star's life. Once the distance to any one group is determined, one can find distances to other groups of stars. Main sequence photometry is effective up to distances of $10^5$ pc. Beyond this distance, it is difficult to find bright enough stars in the main sequence to observe. This method has been discussed in more detail in Chapter 6.

\section{Variable stars}
Variable stars are those whose apparent brightness as seen from the earth changes with time.
The bright Cepheid variables have a typical relation between the luminosity and the period of change in luminosity. \footnote{Henrietta Swan Leavitt discovered this relation while studying Cepheids in the Small Magellanic Cloud.}
RR Lyrae variable stars are characterised by their short periods of brightness fluctuations. These have been observed to have a uniform intrinsic brightness of about fifty times that of the sun on average. Thus, once a Cepheid variable or RR Lyrae has been identified in a cluster of stars, it is possible to estimate its distance as absolute luminosity is known. Cepheids can now be observed from space telescopes at distances up to 30 Mpc which covers the local group of galaxies that the Milky Way is part of.

\section{Other methods}
For greater distances still, say up to $10^{10}$ pc, very very bright objects like galaxies and supernovae are observed. While type \Rmnum{1}a  supernovae are believed to be excellent standard candles, bright galaxies are not thought to be. This is because the pattern in variation in their absolute luminosities over time is not yet clearly understood. There are, however, two relations that help in determining absolute luminosities of galaxies.

The Tully-Fisher relation, published by astronomers R. Brent Tully and J. Richard Fisher, provides a way to estimate the absolute luminosities of spiral galaxies. Tully and Fisher studied the spectra of galaxies to find their maximum rotational velocity. They found that these could be related to the absolute luminosities of the galaxies. Maximum rotational velocity of a galaxy is related to its total mass and hence, to its absolute luminosity. The particular feature of the spectrum usually studied is the widening of the 21 cm absorption line due to transition of states in Hydrogen atoms. The widening of spectral
lines is caused by the motion of stars both towards and away from the observer in a rotating galaxy. The extent to which the lines widen and become less sharp is an indication of the maximum rotational speed of the galaxy.

The Faber-Jackson relation is similar to the Tully-Fisher relation in that it uses the velocities of stars in galaxies to make predictions about its absolute luminosity. However, the random velocities of stars in the centre of a galaxy are used to arrive at a conclusion about the total mass in a galaxy as against rotational velocities used in the Tully-Fisher relation. Greater the mass of a gravitationally bound system like a galaxy, higher the random velocities of individual stars in it need to be in order for the system to be stable. The mass, so estimated, is then related to the absolute luminosity of the galaxy by the relation. This relation works for elliptical galaxies. \footnote{Reference: Steven Weinberg's Cosmology, 2008 and Gravitation and Cosmology, 1972}

\newpage
\pagecolor{black}
\includegraphics[width=160mm]{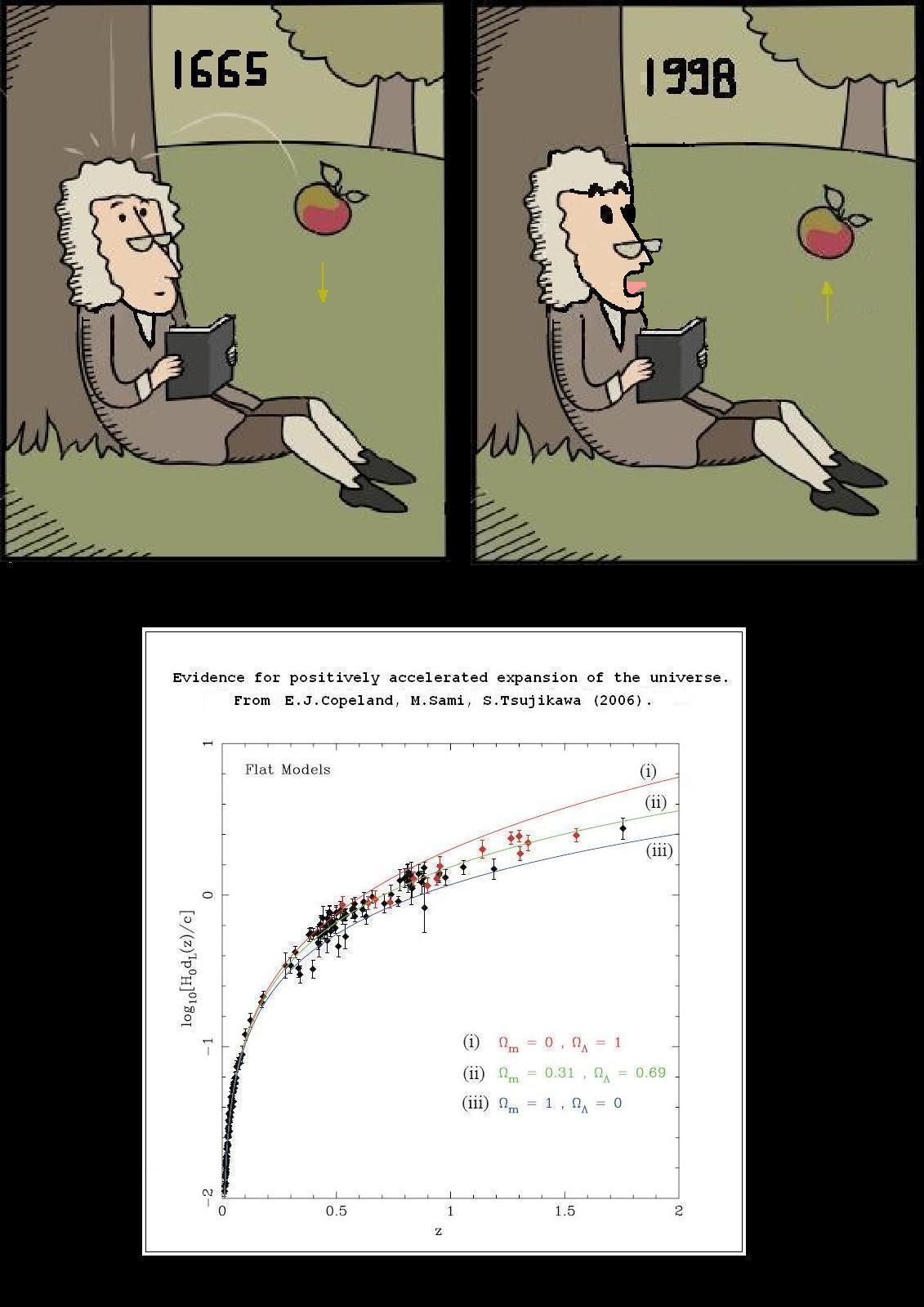}

\begin{thebibliography}{1}
\bibitem{anisotropies} Some of the experiments that have, over the years, probed the anisotropies in the CMBR spectrum include COBE, MAXIMA, BOOMERanG, CBI, ACBAR and WMAP.
\end{thebibliography}
\end{document}